\documentclass[12pt,reprint,aps,groupedaddress,nofootinbib,prd,twocolumn]{revtex4-1}

\usepackage[english]{babel}
\usepackage[utf8]{inputenc}
\usepackage{amsmath}
\usepackage{mathbbol}
\usepackage{amssymb}
\usepackage{bbold}
\usepackage{graphicx,amsfonts}
\usepackage{epsfig}
\usepackage{xcolor}
\definecolor{crimsonglory}{rgb}{0.75,0.0,0.2}
\usepackage[colorlinks=true,
linkcolor=crimsonglory,
urlcolor=crimsonglory,
citecolor=crimsonglory]{hyperref}
\usepackage{capt-of}
\usepackage{bm}
\usepackage{mathrsfs}
\usepackage{enumerate}
\usepackage{amsthm}
\usepackage{bbm}
\usepackage{comment}
\usepackage{appendix}
\usepackage{physics}
\usepackage{url}
\usepackage{upgreek}

\newcommand{\be}{\begin{equation}}
	\newcommand{\ee}{\end{equation}}
\newcommand{\beq}{\begin{equation}}
	\newcommand{\eeq}{\end{equation}}
\newcommand{\bea}{\begin{eqnarray}}
	\newcommand{\eea}{\end{eqnarray}}
\newcommand{\bit}{\begin{itemize}}
	\newcommand{\eit}{\end{itemize}}
\newcommand{\ben}{\begin{enumerate}}
	\newcommand{\een}{\end{enumerate}}

\begin{document}
	
	\preprint{APS/123-QED}
	
	\title{Hydrogen-like structures in the strong interaction}
	
	\author{Lei Liu}
	\affiliation{College of Physics, Chengdu University of Technology, Chengdu 610059, China}
	
	\author{Yanmei Xiao}
	\affiliation{College of Physics, Chengdu University of Technology, Chengdu 610059, China}
	
	\author{Tao Guo}
	\email[Corresponding author:~]{guot17@tsinghua.org.cn}
	\affiliation{College of Physics, Chengdu University of Technology, Chengdu 610059, China}

	\date{\today}%
	
\begin{abstract}
		
		Heavy flavor hadrons, especially doubly heavy baryons and doubly heavy tetraquarks, have always received extensive attention in theoretical and experimental research.
		Given the separation of quark masses $m_Q \gg m_q$ ($Q = c, b$ and $q = u, d, s$), this type of heavy flavor hadrons can be well regarded as hydrogen-like structures in the strong interaction.
		In the theoretical framework of Born-Oppenheimer approximation, we derive the Schr{\"o}dinger equation for the motion of light quarks in the effective potential field of heavy quarks.
		Taking proper account of the color-spin hyperfine interaction, we carry out a systematic study on the mass spectra of $S$-wave doubly heavy baryons and doubly heavy tetraquarks.
		The model parameters required for the calculation are obtained by fitting conventional hadrons.
		The investigation on the doubly heavy baryon systems indicates the reliability of our theoretical approach.
		Our calculation results show that the experimentally discovered $T_{cc}^+$ is most likely a compact tetraquark $|(cc)_0^{6}(\bar{u}\bar{d})_1^{\bar{6}}\rangle$ state with quantum numbers $(I,J^P)=(0, 1^+)$.
		Forthermore, for different quantum number assignments, some stable tetraquark states are found and may be very narrow peaks.
		The predictions for other heavy flavor hadrons are expected to be confirmed in new theories and future experiments.
		
\end{abstract}
	
	\maketitle
	
\section{Introduction}
	
\begin{figure}[h]
	\centering
	\includegraphics[width=0.80\linewidth]{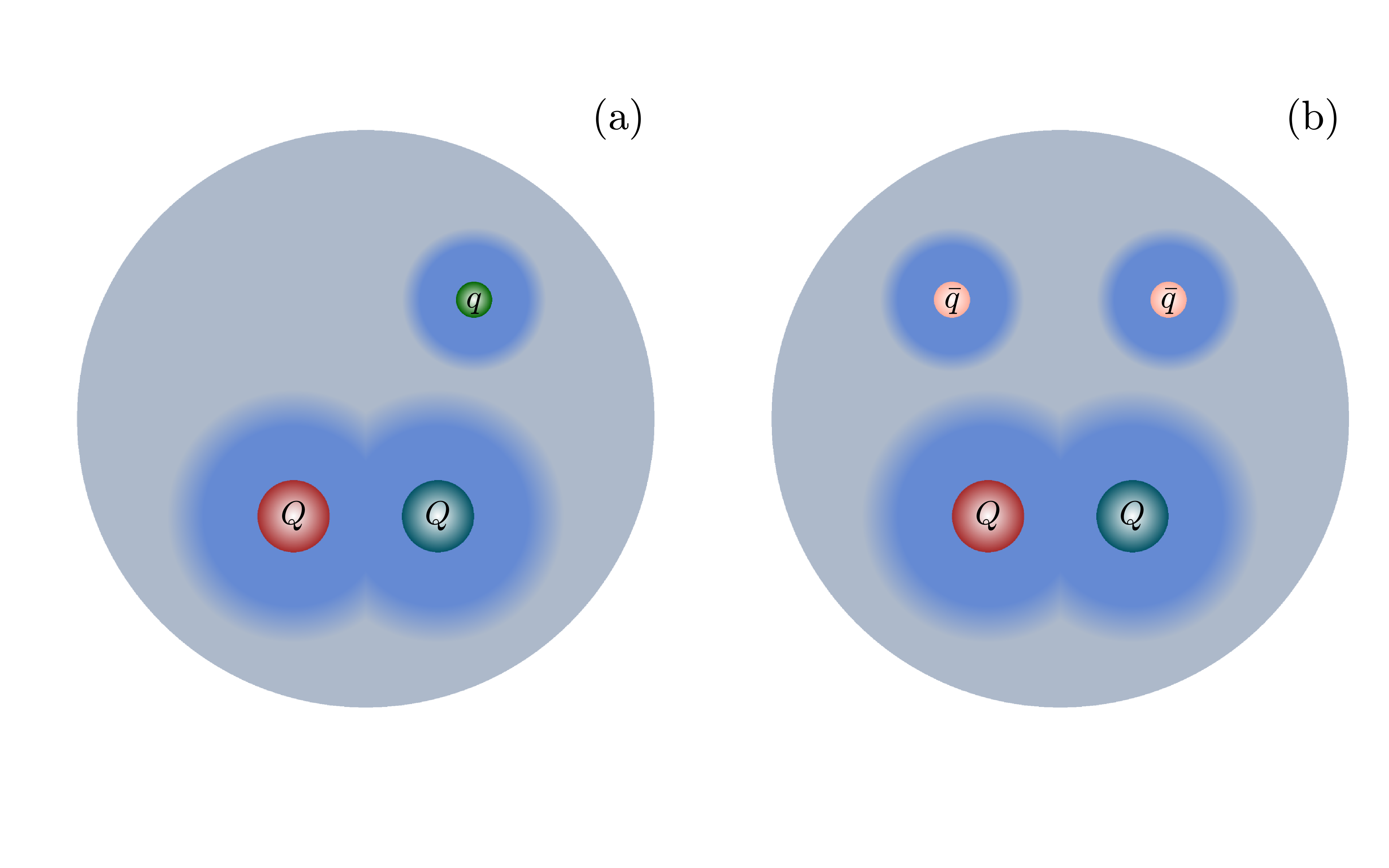}
	\vspace{-0.6cm}
	\caption{Schematic diagram of a doubly heavy baryon (a) and a doubly heavy tetraquark (b).}
	\label{figbt}
\end{figure}

For a long time, one of the core tasks of particle physics has been to reveal the microscopic structure of hadrons and the laws of interaction between the basic units (quarks and gluons) that constitute them.
Within this framework, the quark model \cite{Gell-Mann:1964ewy} and the quantum chromodynamics (QCD) \cite{Fritzsch:1973pi,Gross:1973id} describing the strong interaction came into being.
However, due to the self-interaction of gluons, QCD theory cannot be used well to deal with physical problems on the low-energy hadronic scale.
A series of phenomenological models and numerical methods based on the QCD first principle have been developed, such as the effective potential model, QCD sum rule, and lattice QCD calculation, see the reviews \cite{Klempt:2007cp,Liu:2019zoy,Brambilla:2019esw} for details.
At the same time, with the discovery of more new hadrons in experiments \cite{ParticleDataGroup:2024cfk}, it is very challenging to further improve the precision and accuracy of existing theoretical calculations.
In the study of hadronic physics, the heavy flavor hadrons composed of heavy and light quarks, especially doubly heavy baryons ($QQq$) and doubly heavy tetraquarks ($QQ\bar{q}\bar{q}$), have aroused our great interest, as shown in Figure \ref{figbt}.
Based on the fact that heavy quarks are much heavier than light quarks, such doubly heavy flavor hadrons can be naturally regarded as hydrogen-like structures in the strong interaction.

In a hydrogen molecule ($\text{H}_2$), the Born–Oppenheimer (BO) approximation assumes that the wave functions of protons and electrons can be treated separately.
This means that the energy of hydrogen molecule can be expressed as the sum of several independent terms.
The dissociation energy of hydrogen molecule calculated using the BO approximation is in excellent agreement with the experimental value, and the accuracy reaches the $10^{-8}$ level \cite{Piszczatowski2009,Liu2009}.
Similarly, the BO approximation theory can be used to calculate hydrogen-like structures in the strong interaction, except that the original Coulomb interactions need to be replaced by the color Coulomb interactions.
Currently, this research paradigm is being rapidly promoted and used in the field of hadronic physics, specifically involving heavy flavor baryons \cite{Maiani:2019lpu}, multiquark states \cite{Braaten:2014qka,Bicudo:2017szl,Maiani:2019cwl,Giron:2019bcs,Bruschini:2020voj,Germani:2020sgm,Mutuk:2024vzv,Kang:2025xqm}, and hybrids \cite{Berwein:2024ztx}.
Furthermore, an effective field theory based on the BO approximation is established, allowing the calculation of the next-to-leading order corrections to the multipole expansion of the heavy quark potential in QCD molecules by a systematic and controllable manner \cite{Brambilla:2017uyf}.
With the above cases as a guide, the BO approximation can be one of the effective tools for studying doubly heavy baryons and doubly heavy tetraquarks.

The study of heavy flavor hadron systems has always attracted keen attention in both theoretical and experimental fields.
In 2003, the SELEX Collaboration experimentally observed for the first time the doubly charmed baryon $\Xi_{cc}^+(3519)$, with a statistical significance of 6.3$\sigma$ \cite{SELEX:2002wqn}.
Unfortunately, the subsequent FOCUS, BaBar, Belle, and LHCb experiments did not reproduce the observational evidence for $\Xi_{cc}^+$ \cite{Ratti:2003ez,BaBar:2006bab,Belle:2006edu,LHCb:2013hvt}.
In 2017, another doubly charmed baryon $\Xi_{cc}^{++}(3621)$ was observed by the LHCb collaboration in the $\Lambda_c^+ K^-\pi^+\pi^-$ invariant mass spectrum \cite{LHCb:2017iph}.
And the $\Xi_{cc}^{++}$ was soon further confirmed in the decay modes $\Xi_{cc}^{++} \rightarrow \Xi_{c}^{+}\pi^+$ \cite{LHCb:2018pcs} and $\Xi_{cc}^{++} \rightarrow \Xi_{c}^{\prime +}\pi^+$ \cite{LHCb:2022rpd}.
Recently, the LHCb collaboration observed a narrow peak $T_{cc}^+(3875)$ in the $D^0D^0\pi^+$ invariant mass spectrum, whose mass is very close to and slightly below the $D^{*+}D^0$ mass threshold \cite{LHCb:2021vvq,LHCb:2021auc}.
The quark content and spin-parity quantum numbers of the $T_{cc}^+$ are $cc\bar{u}\bar{d}$ and $J^P = 1^+$ respectively, which is in good agreement with the theoretical expectations about tetraquark state \cite{Navarra:2007yw,Ohkoda:2012hv,Ikeda:2013vwa,Karliner:2017qjm,Guo:2021yws,Lyu:2023xro,Meng:2024kkp}.
Of course, there are also some other theoretical calculations suggesting that $T_{cc}^+$ may be a molecular state \cite{Fleming:2021wmk,Ling:2021bir,Du:2021zzh,Gil-Dominguez:2024zmr} or a threshold effect \cite{Braaten:2022elw,Achasov:2022onn}.
In this work, we employ the BO approximation to systematically study the $S$-wave double heavy hadrons ($QQq$ and $QQ\bar{q}\bar{q}$) and predict several possible bound states. 
With proper consideration of the color Coulomb interaction and the color-spin hyperfine interaction, this method provides a new perspective and a reliable theoretical calculation framework for the study of this type of hadrons.

This paper is organized as follows. 
In Sec. \ref{S2}, we set up the Hamiltonian for doubly heavy hadron systems in the BO approximation.
We then give a brief introduction to solving the BO potential and the basis wave functions of $S$-wave double heavy baryons and double heavy tetraquarks.
Besides, the model parameters are also extracted in this section.
In Sec. \ref{S3}, we systematically calculate the mass spectra of $S$-wave double heavy baryons and discuss as well as analyze the application of the BO approximation to these systems. 
The mass spectra and decay channels of $S$-wave doubly heavy tetraquarks are calculated and discussed in Sec. \ref{S4}. 
In the last section, we give a summary of this work.

\section{Theoretical approach \label{S2}}

The color confinement and gauge symmetry of QCD requires that any hadron composed of quarks and gluons must be a color singlet.
For the $S$-wave hadrons, the spin-orbit angular momentum coupling term is set to zero by default.
Here, we introduce the BO approximation theoretical framework to study doubly heavy baryons and doubly heavy tetraquarks.
The Hamiltonian of the system can be expressed as \cite{Kang:2025xqm}:
\begin{equation}
	H  = \sum_{i=1}^N m_i+E+H_{ss},
	\label{H}
\end{equation}
where the color-spin hyperfine interaction term is
\begin{equation}
	H_{ss} = \sum_{i<j} 2\kappa_{ij} (\bm{s}_i \cdot \bm{s}_j).
	\label{Hss}
\end{equation}
In Eq.(\ref{H}), $N$ is the number of constituent quarks in a multiquark hadron.
$m_i$ represents the effective mass of the $i$-th constituent quark.
$E$ is the total ground-state energy eigenvalue of the system calculated based on the BO approximation.
In Eq.(\ref{Hss}), the model parameter $\kappa_{ij}$ is related to the masses of the constituent quarks, the color coupling strength, the running coupling constant, and the spatial configuration of the hadron. 
Without loss of generality, the above parameter $\kappa_{ij}$ can be extracted by fitting the conventional hadrons. 
$\bm{s}_i$ represents the spin operator of the $i$-th constituent quark.

\subsection{Born-Oppenheimer approximation}

To study doubly heavy baryons and doubly heavy tetraquarks, we adopt the BO approximation.
In such a hadronic system, the masses of the heavy quarks are significantly larger than those of light quarks. 
Given the same momentum, this means that the velocities of heavy quarks are much smaller than those of light quarks. 
The BO approximation assumes that the wave functions of the heavy and light quarks can be treated separately.
In our calculations, the coordinates of the heavy quarks are denoted as $\bm{x_A}$ or $\bm{x_B}$, while the coordinates of the light quarks are denoted as $\bm{x_1}$ or $\bm{x_2}$. 
For a heavy flavor hadron consisting of heavy and light quarks, the total nonrelativistic Hamiltonian reads
\begin{equation}
		H_{{t}}= H_{{h}}+H_{{l}},
	\label{H_total}
\end{equation}
where the Hamiltonian for heavy quarks is
\begin{equation}
		H_{{h}}=\sum_{\text{heavy}}\frac{p_Q^2}{2m_Q}+V(\bm{\bm{x_A}},\bm{x_B}),
	\label{H_h}
\end{equation}
and the Hamiltonian for light quarks is
\begin{equation}
		H_{{l}}=\sum_{\text{light}}\frac{p_q^2}{2m_q}+V(\bm{x_A},\bm{x_B};\bm{x_1},\bm{x_2}).
	\label{H_l}
\end{equation}
Here, $p_Q$ ($p_q$) and $m_Q$ ($m_q$) represent the momentum and mass of heavy (light) quark respectively.
The interaction term between two quarks is decomposed into two parts: $V(\bm{x_A}, \bm{x_B})$ represents the heavy-heavy quark interaction, and $V(\bm{x_A}, \bm{x_B}; \bm{x_1}, \bm{x_2})$ represents the heavy-light as well as light-light quark interactions. 
Assume that the total wave function of the system can be written as
\begin{equation}
	\Psi = \psi(\bm{x_A}, \bm{x_B}) f(\bm{x_A}, \bm{x_B}; \bm{x_1}, \bm{x_2}),
	\label{Psi}
\end{equation}
where $\psi=\psi(\bm{x_A}, \bm{x_B})$ describes the wave function of the heavy quarks, and $f=f(\bm{x_A}, \bm{x_B}; \bm{x_1}, \bm{x_2})$ describes the wave function of the associated light quarks.

In the BO approximation, substituting Eq.(\ref{Psi}) into Eq.(\ref{H_total}), we can obtain the total Schr{\"o}dinger equation for the system as
\begin{equation}
	\left( \sum_{\text{heavy}}\frac{p_Q^2}{2m_Q}+V_{\text{BO}}(\bm{x_A},\bm{x_B})\right)\psi=E\psi,
	\label{heavy_S}
\end{equation}
with BO potential
\begin{equation}
	V_{\text{BO}}(\bm{x_A},\bm{x_B})=E_{{l}}(\bm{x_A},\bm{x_B}) + V(\bm{x_A},\bm{x_B}).
	\label{BO_potentional}
\end{equation}
In the above Eq.(\ref{BO_potentional}), $E_l = E_{l}(\bm{x_A},\bm{x_B})$ is the energy eigenvalue of the light quarks in all fixed heavy quark potential fields.
In general, it can be obtained by solving the Schr{\"o}dinger equation
\begin{equation}
	H_{{l}}f=E_{{l}}f.
	\label{light_S}
\end{equation}
Alternatively, $V(\bm{x_A},\bm{x_B})$ represents the color Coulomb interaction between two heavy flavor quarks, which originates from the one-gluon-exchange process.
However, under existing conditions, the color Coulomb interaction alone is not enough to allow multiple quarks to combine into a color-singlet hadron.
In a baryon or tetraquark system, the confinement is known to occur.
Thus, we introduce an additional linear increase term $V_{\text{conf}}$ to simulate the confinement feature.
And it is determined by the string tensor strength $k$ and the adjustable starting point $R_0$ \cite{Andreev:2022cax}.
The parameter $R_0$ is related to the spatial configuration of the hadron system.
Then the heavy-heavy quark interaction term is written as
\begin{equation}
 V(\bm{x_A},\bm{x_B})=\lambda_{Q_1Q_2}\frac{\alpha_s}{r_{AB}}+V_{\text{conf}},
	\label{h-h}
\end{equation}
with the linear confinement term
\begin{equation}
	V_{\text{conf}}=k \times (r_{AB}-R_0) \times \theta(r_{AB}-R_0),
	\label{conf}
\end{equation}
where $\lambda_{Q_1Q_2}$ is the quadratic Casimir operator for a quark pair in the color $SU(3)$ group representation.
$\alpha_s$ is the running coupling constant of QCD, and $r_{AB}= |\bm{r_{AB}}| = |\bm{x_{B}}-\bm{x_{A}}|$ is the relative distance between the two heavy flavor quarks.

\subsection{Color-spin wave function}
In a hadron system, hyperfine interaction induces splitting of energy levels.
From Eq.(\ref{Hss}), we can see that the hyperfine interaction is related to the color and spin of the quark pairs.
Now we construct the color-spin wave functions of the doubly heavy baryons and doubly heavy tetraquarks, respectively.

\textit{Doubly heavy baryons}. 
For an $S$-wave doubly heavy baryon $Q_1Q_2q$, the total spin quantum numbers can be assigned to $\frac{1}{2}$ and $\frac{3}{2}$.
The color coupling between any two quarks can only be $\bar{3}$, and then coupled with the third quark to form a color singlet, i.e., $\bar{3}\otimes 3 \rightarrow 1$.
Using the diquark-quark configuration, we construct the color-spin basis vectors as follows.

For the quantum number $J^P=\frac{1}{2}^+$, the color-spin basis vectors read
\begin{equation}
	\begin{aligned}
		&\mu_1=|(Q_1Q_2)_0^{\bar{3}}(q)_{\frac{1}{2}}^{3}\rangle,\\
        &\mu_2=|(Q_1Q_2)_1^{\bar{3}}(q)_{\frac{1}{2}}^{3}\rangle.
        \label{b12}
	\end{aligned}
\end{equation}
In the above basis vectors, the superscripts indicate the color and the subscripts indicate the spin. 
This notation is used throughout the paper unless otherwise stated.
It is worth noting that in such a baryon system, the Pauli exclusion principle of fermions should be satisfied.
When the two heavy quarks that constitute the doubly heavy baryon are of the same flavor, the color-spin basis vector $\mu_1$ does not exist. 
Thus, there is only one color-spin basis vector $\mu_2$ in this system.

For $J^P=\frac{3}{2}^+$, the color-spin basis vector is
\begin{equation}
	   \nu_1=|(Q_1Q_2)_1^{\bar{3}}(q)_{\frac{1}{2}}^{3}\rangle.
	   \label{b32}
\end{equation}
Next, we only need to substitute Eqs.(\ref{b12}) and (\ref{b32}) into the hyperfine interaction Hamiltonian (\ref{Hss}) and calculate the expectation values to obtain the energy level splitting of the systems.
Furthermore, we can solve the mass spectra and wave functions of all $S$-wave doubly heavy baryon systems.

\textit{Doubly heavy tetraquarks}. 
Physically, the configurations of the doubly heavy tetraquark $Q_1Q_2\bar{q}_3\bar{q}_4$ usually include two types: diquark-antidiquark ($ |(Q_1Q_2)(\bar{q}_3\bar{q}_4)\rangle $) and meson-meson ($|(Q_1\bar{q}_3)(Q_2\bar{q}_4)\rangle$ or $|(Q_1\bar{q}_4)(Q_2\bar{q}_3)\rangle$).
In specific calculations, these two types of configurations are connected by Fierz identity.
The total wave function of the system can be written as the direct product of the flavor, spin, color, and orbital wave functions.
For the $S$-wave doubly heavy tetraquarks, the orbital wave function is symmetric, and the total spin quantum numbers can be 0, 1, and 2.
In order to consider the identical principle of fermions, we choose the diquark-antidiquark configurations in this paper.

In the color space, the color couplings of diquark and antidiquark are $3\otimes3 \rightarrow \bar{3} \oplus 6$ and $\bar{3}\otimes\bar{3} \rightarrow 3 \oplus \bar{6}$, respectively.
Here we do not consider the mixing effect of states $|(Q_1Q_2)^{\bar{3}}(\bar{q}_3\bar{q}_4)^{3}\rangle$ and $|(Q_1Q_2)^6(\bar{q}_3\bar{q}_4)^{\bar{6}}\rangle$. 
The advantage is that we can quantitatively give the mass spectra and wave functions of the tetraquark systems in specific color coupling.
Taking proper account of the symmetry, the color-spin basis vectors of the $S$-wave doubly heavy tetraquarks are constructed.

For the $|(Q_1Q_2)^{\bar{3}}(\bar{q}_3\bar{q}_4)^{3}\rangle$ configuration, the color-spin basis vectors of the system in different quantum numbers can be built as
\begin{enumerate}
	\item $J^P=0^+$
	\begin{equation}
		\begin{aligned}
			&\alpha_1=|(Q_1Q_2)_0^{\bar{3}}(\bar{q}_3\bar{q}_4)_0^{3}\rangle\delta_{12}^S\delta_{34}^S,\\
			&\alpha_2=|(Q_1Q_2)_1^{\bar{3}}(\bar{q}_3\bar{q}_4)_1^{3}\rangle\delta_{12}^A\delta_{34}^A.\\
		\end{aligned}
	\end{equation}
	\item $J^P=1^+$
	\begin{equation}
		\begin{aligned}
			&\beta_1=|(Q_1Q_2)_0^{\bar{3}}(\bar{q}_3\bar{q}_4)_1^{3}\rangle\delta_{12}^S\delta_{34}^A,\\
			&\beta_2=|(Q_1Q_2)_1^{\bar{3}}(\bar{q}_3\bar{q}_4)_0^{3}\rangle\delta_{12}^A\delta_{34}^S,\\
			&\beta_3=|(Q_1Q_2)_1^{\bar{3}}(\bar{q}_3\bar{q}_4)_1^{3}\rangle\delta_{12}^A\delta_{34}^A.\\
		\end{aligned}
	\end{equation}
	\item $J^P=2^+$
	\begin{equation}
		\begin{aligned}
			&\gamma_1=|(Q_1Q_2)_1^{\bar{3}}(\bar{q}_3\bar{q}_4)_1^{3}\rangle\delta_{12}^A\delta_{34}^A.\\
		\end{aligned}
	\end{equation}
\end{enumerate}

Similarly, the color-spin basis vectors in the tetraquark $|(Q_1Q_2)^6(\bar{q}_3\bar{q}_4)^{\bar{6}}\rangle$ configuration are expressed as
\begin{enumerate}
	\item $J^P=0^+$
	\begin{equation}
		\begin{aligned}
			&\alpha_3=|(Q_1Q_2)_0^{6}(\bar{q}_3\bar{q}_4)_0^{\bar{6}}\rangle\delta_{12}^A\delta_{34}^A,\\
			&\alpha_4=|(Q_1Q_2)_1^{6}(\bar{q}_3\bar{q}_4)_1^{\bar{6}}\rangle\delta_{12}^S\delta_{34}^S.
		\end{aligned}
	\end{equation}
	\item $J^P=1^+$
	\begin{equation}
		\begin{aligned}
			&\beta_4=|(Q_1Q_2)_0^{6}(\bar{q}_3\bar{q}_4)_1^{\bar{6}}\rangle\delta_{12}^A\delta_{34}^S,\\
			&\beta_5=|(Q_1Q_2)_1^{6}(\bar{q}_3\bar{q}_4)_0^{\bar{6}}\rangle\delta_{12}^S\delta_{34}^A,\\
			&\beta_6=|(Q_1Q_2)_1^{6}(\bar{q}_3\bar{q}_4)_1^{\bar{6}}\rangle\delta_{12}^S\delta_{34}^S.
		\end{aligned}
	\end{equation}
	\item $J^P=2^+$
	\begin{equation}
		\begin{aligned}
			&\gamma_2=|(Q_1Q_2)_1^{6}(\bar{q}_3\bar{q}_4)_1^{\bar{6}}\rangle\delta_{12}^S\delta_{34}^S.
		\end{aligned}
	\end{equation}
\end{enumerate}
In the above basis vectors, we introduce the symbols $\delta_{ij}^A$ and $\delta_{ij}^S$ to satisfy the exchange symmetry principle of fermions.
If the two heavy quarks in the system are of the same flavor, then $\delta_{12}^S=0$ and $\delta_{12}^A=1$.
If the two heavy quarks are different, we have $\delta_{12}^S=\delta_{12}^A=1$.
Considering that the two light antiquarks in the system are $\bar{u}$ or $\bar{d}$, the doubly heavy tetraquark system has a definite isospin.
For the isospin $I=0$, we have $\delta_{34}^S=1$ and $\delta_{34}^A=0$. 
For the isospin $I=1$, we have $\delta_{34}^S=0$ and $\delta_{34}^A=1$.
In addition, if the light antiquark pair $\bar{q}_3\bar{q}_4$ is $\bar{s}\bar{s}$, its symmetry satisfies the isospin $I=1$ case.
If there is only one $\bar{s}$ in $\bar{q}_3\bar{q}_4$, we take $\delta_{34}^S=\delta_{34}^A=1$.

\subsection{Model parameters}

In order to calculate the mass spectra of doubly heavy baryons and doubly heavy tetraquarks, we first need to determine the values of the model parameters.
Considering the flavor symmetry breaking of light quarks, we introduce the symbol $n=u, d$ to distinguish $s$ quark.
The effective masses of the constituent quarks can be obtained by fitting the experimental values of $S$-wave mesons and baryons, i.e.,
$m_n = 308$ MeV, $m_s = 484$ MeV, $m_c = 1667$ MeV, and $m_b = 5005$ MeV.
An effective scale-dependent coupling constant reads \cite{Vijande:2004he}:
\begin{equation}
\alpha_s\left(\mu\right)=\frac{\alpha_0}{\text{ln}\left(\frac{\mu^2+\mu_0^2}{\Lambda_0^2}\right)},
\end{equation}
where the model parameters are denoted as: $\alpha_0 = 2.118$, $\Lambda_0 = 0.113$ fm$^{-1}$, and $\mu_0 = 36.976 $ MeV.
The symbol $\mu$ is the reduced mass of the quark-quark or quark-antiquark pair.
The string tensor strength $k = 0.15$ GeV$^2$ is extracted by fitting the masses of charmonium, bottomonium, and charm-bottom mesons.

	\begin{table}[h]
		\centering
		\caption{ Parameters $(\kappa_{ij})^1$ and $(\kappa_{ij})^{\bar{3}}$ (in MeV) correspond to quark-antiquark and quark-quark systems respectively.}
		\label{table1}
		\begin{ruledtabular}
			\begin{tabular}{ccccccccccc} 
				Mesons & $n\bar{n}$ & $n\bar{s}$ & $s\bar{s}$ & $n\bar{c}$ & $s\bar{c}$ & $c\bar{c}$ & $n\bar{b}$ & $s\bar{b}$ & $c\bar{b}$ & $b\bar{b}$\\ 
				\hline
				$(\kappa_{ij})^1$ & $315$ & $121$ & $195$ & $70$ & $72$ & $59$ & $23$ & $24$ & $20$ & $30$\\ 
				
				Baryons & $nn$ & $ns$ & $ss$ & $nc$ & $sc$ & $cc$ & $nb$ & $sb$ & $cb$ & $bb$ \\ 
				
				$(\kappa_{ij})^{\bar{3}}$ & $103$ & $64$ & $72$ & $22$ & $25$ & $7.4$ & $6.6$ & $7.5$ & $10$ & $3.9$\\ 
				
			\end{tabular}
		\end{ruledtabular}
	\end{table}

Finally, the parameters required to calculate the color-spin hyperfine interaction are listed in Table \ref{table1}.
As mentioned above, the superscripts of parameters $(\kappa_{ij})^1$ and $(\kappa_{ij})^{\bar{3}}$ also indicate the colors of the quark pairs.
The extraction results of most parameters in Table \ref{table1} have been given in the Ref.\cite{Kang:2025xqm}.
The calculation strategies for the remaining parameters are similar and will not be described in detail here.
Furthermore, considering the symmetry, these parameters satisfy $\kappa_{ij} = \kappa_{ji}$.

\section{Doubly heavy baryons \label{S3}}

In this section, we systematically investigate the $S$-wave states of doubly heavy baryons based on the BO approximation.
The specific calculation framework is proposed based on the similarity between doubly heavy baryon $Q_1Q_2q$ and hydrogen molecular ion $\text{H}_2^+$.
The physical parameters and color-spin basis wave functions required for the calculation have been given in the previous section.
By solving the expectation value of the Hamiltonian (\ref{H}), we can obtain the mass spectra of $S$-wave doubly heavy baryons with quantum numbers $J^P=\frac{1}{2}^+$ and $\frac{3}{2}^+$.
For the doubly heavy baryon systems, the universal interaction term is written as
\begin{equation}
		V_t=V(\bm{x_A},\bm{x_B})+V(\bm{x_A},\bm{x_B};\bm x),
\end{equation}
with the heavy-heavy quark interaction
\begin{equation}
	V(\bm{x_A},\bm{x_B})=-\frac{2}{3}\frac{\alpha_s}{r_{AB}} + V_{\text{conf}},
\end{equation}
and the heavy-light quark interaction
\begin{equation}
	V(\bm{x_A},\bm{x_B};\bm x)=-\frac{2}{3}\frac{\alpha_s^\prime}{r_A}-\frac{2}{3}\frac{\alpha_s^\prime}{r_B}.
\end{equation}
Here, the distances between a light quark and two heavy quarks are represented as $r_A = |\bm x - \bm{x_A}|$ and $r_B = |\bm x - \bm{x_B}|$ respectively.
$\alpha_s^\prime$ is the strong coupling constant between light and heavy quarks, where the mass of the heavy diquark is replaced by the reduced mass.

Now we solve the Schr{\"o}dinger Eq.(\ref{light_S}) for light quark in the heavy quark potential field. 
The Hamiltonian for light quark is simplified to
\begin{equation}
	H_{l}=-\frac{1}{2m_q}\nabla^2-\frac{2}{3}\frac{\alpha_s^\prime}{r_A}-\frac{2}{3}\frac{\alpha_s^\prime}{r_B}.
\end{equation}
Assume that the radial variation wave function of the system reads
\begin{equation}
	R(r)=\frac{A^\frac{3}{2}}{\sqrt{\pi}}e^{-Ar},
\end{equation}
where $A$ is the variational parameter.
For a doubly heavy baryon system, a light quark is affected by the potential fields of two heavy quarks. 
Therefore, the wave function of the light quark can be written as
\begin{equation}
		f=\frac{R(r_A)+R(r_B)}{\sqrt{2(1+\mathcal{J})}} ,
\end{equation}
with the overlap function
\begin{equation}
\mathcal{J}=\int R(r_A)R(r_B)d\tau.
\end{equation}
Through analytical calculations, we can obtain the energy eigenvalue of light quark
\begin{equation}
		E_l=\frac{\langle R(r_A),H_{l}R(r_A)\rangle+\langle R(r_B),H_{l}R(r_A)\rangle}{1+\mathcal{J}}, 
	\label{E_lb}
\end{equation}
with 
\begin{equation}
	\begin{aligned}
		&\langle R(r_B),H_{l}R(r_A)\rangle= (\frac{A}{m_{q}}-\frac{4}{3}\alpha_s^\prime)\mathcal{I}-\frac{A^2}{2m_{q}}\mathcal{J} ,\\
		&\langle R(r_A),H_{l}R(r_A)\rangle= \frac{A^2}{2m_{q}}-\frac{2A}{3}\alpha_s^\prime-\frac{2}{3}\alpha_s^\prime\mathcal{K}.
	\end{aligned}
\end{equation}
The above expectation values include integrals related to the orbital wave functions:
\begin{equation}
	\begin{aligned}
		&\mathcal{I}=\int\frac{R(r_A)R(r_B)}{r_A}d\tau=\int\frac{R(r_A)R(r_B)}{r_B}d\tau ,\\
		&\mathcal{K}=\int \frac{R^2(r_A)}{r_B}d\tau=\int \frac{R^2(r_B)}{r_A}d\tau .
	\end{aligned}
\end{equation}

\begin{figure}[h]
	\centering
	\includegraphics[width=0.85\linewidth]{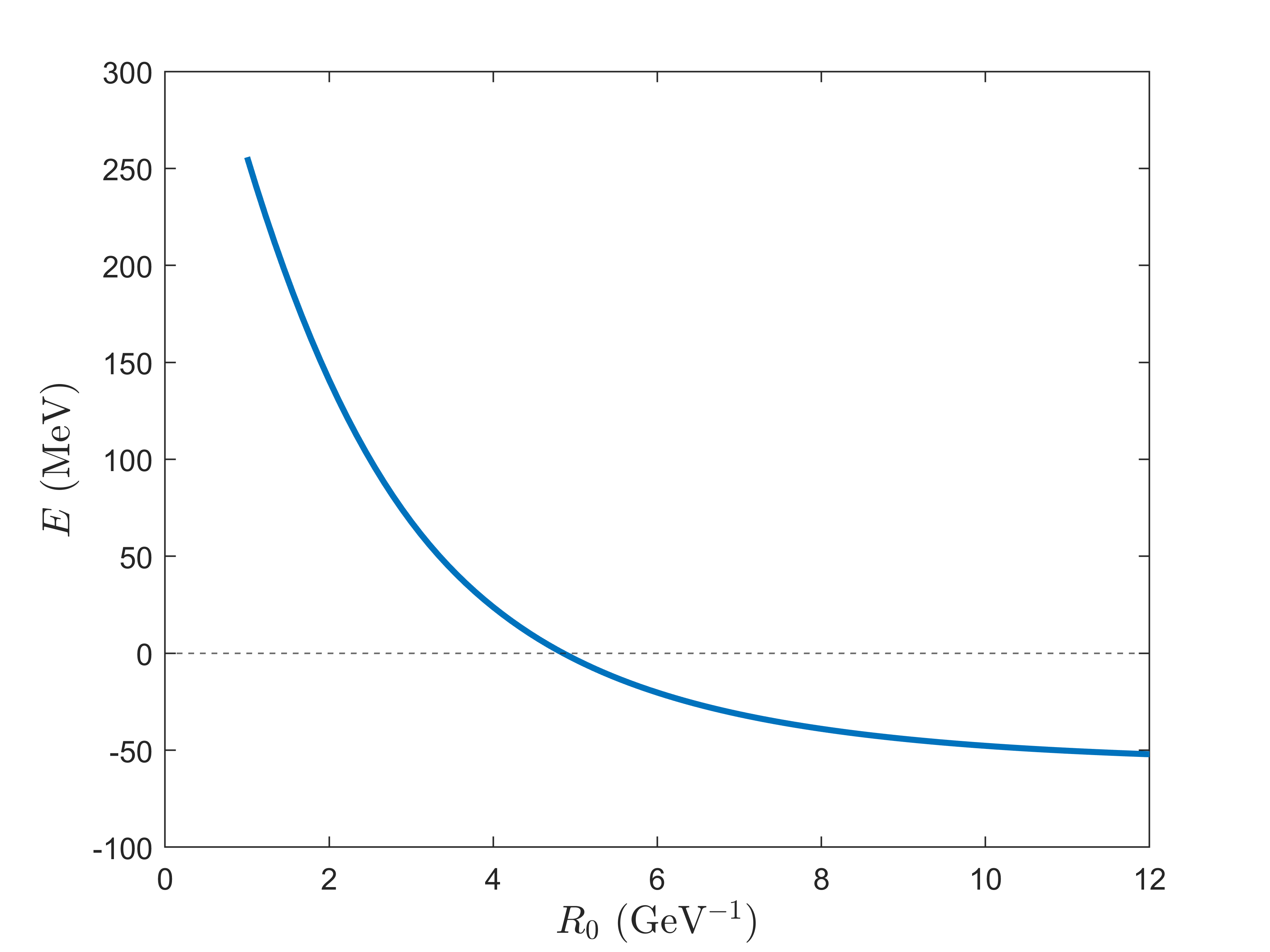}
	\caption{For the doubly charmed baryon $\Xi_{cc}$, the ground-state energy $E$ varies with the parameter $R_0$.}
	\label{fig:E_ccq}
\end{figure}

Using the variational principle on Eq.(\ref{E_lb}), we can obtain the variational parameter $A$ and the minimized energy eigenvalue $E_l$. 
Then substituting $E_l$ into Eq.(\ref{BO_potentional}), the BO potential is expressed as
\begin{equation}
	V_{\text{BO}}(r_{AB})= -\frac{2}{3}\frac{\alpha_s}{r_{AB}}+V_{\text{conf}}+E_l.
	\label{Bo_pb}
\end{equation}
Substituting Eq.(\ref{Bo_pb}) into the Schr{\"o}dinger Eq.(\ref{heavy_S}), we can numerically solve the ground-state energy $E$ of the doubly heavy baryon systems.
Our calculation results show that the ground-state energy $E$ is a function of $R_0$. 
In order to facilitate the calculation of the mass spectra of all doubly heavy baryons, we use the doubly charmed baryon $\Xi_{cc}$ as input to determine the parameter $R_0$.
For the doubly charmed baryon $\Xi_{cc}$, the variation of the total ground-state energy $E$ with parameter $R_0$ is shown in Figure \ref{fig:E_ccq}.
It can be found that $E$ decreases as $R_0$ increases, and eventually tends to be stable.

For the doubly charmed baryon $\Xi_{cc}^{++}$, its quark content is $ccu$. 
Then, the mass spectrum of $\Xi_{cc}^{++}$ with quantum number $J^P=\frac{1}{2}^+$ can be obtained by solving the expectation value of Eq.(\ref{H}), i.e.,
\begin{equation}
	M(\Xi_{cc}^{++})_{\frac{1}{2}^+}=2m_c+m_n+E+\frac{1}{2}(\kappa_{cc})^{\bar{3}}-2(\kappa_{nc})^{\bar{3}}.
\end{equation}
By fitting the experimental value of the doubly charmed baryon $\Xi_{cc}^{++}$, we can obtain the ground-state energy $E$ and the parameter $R_0$ as follows:
\begin{equation}
	E=19.7~\text{MeV},~~R_0 = 4.12 ~\text{GeV}^{-1}.
\end{equation}
It can be found that the ground-state energy of the baryon $\Xi_{cc}^{++}$ is slightly greater than zero, indicating that the three quarks $ccu$ cannot form a bound hadron.
In our calculations, this result may be caused by the difference in the extracted model parameters.
Since $|E/M(\Xi_{cc}^{++})|\sim 0.005$, it can be considered that our calculation results are within the allowable error range.

Using a similar strategy, we can calculate the ground-state energies of other $S$-wave doubly heavy baryon systems based on the BO approximation.
Given the parameter $R_0 = 4.12$ GeV$^{-1}$, the calculation results are obtained and shown in Table \ref{dhbe}.
It can be concluded that the total energy eigenvalues of the doubly heavy baryons containing $b$ quark tend to be more negative.
This indicates that, for a fixed $R_0$, the heavier the doubly heavy baryon, the lower its ground-state energy. 
The main reason is that the color coupling of any diquark in a baryon can only be $\bar{3}$. 
Then the diquark and the third quark form a color singlet baryon from the color coupling $3\otimes\bar{3}\rightarrow 1$. 
In this process, the color Coulomb interaction between any diquark is attractive.

\begin{table}[httb]
	\centering
	\begin{ruledtabular}
		\caption{In the BO approximation, the ground-state energies $E$ (in MeV) of other doubly heavy baryons. In this Table, we take the parameter $R_0 = 4.12$ GeV$^{-1}$.}
		\label{dhbe}
		\begin{tabular}{cccccc}
			
			baryons & $ccs$ & $bbn$ & $bbs$  & $cbn$ & $cbs$ \\
			\hline
			$E$  & 12.4 & -54.3 & -60.5 & -13.7 & -20.4 \\
			
		\end{tabular}
	\end{ruledtabular}
\end{table}

Further considering the color-spin hyperfine interaction, we can obtain the mass spectra of the $S$-wave doubly heavy baryons with quantum numbers $J^P=\frac{1}{2}^+$ and $\frac{3}{2}^+$.
The results are shown in Tables \ref{baryons_mass1}-\ref{baryons_mass2}.
At the same time, the results of some other theoretical calculations are also listed in Tables \ref{baryons_mass1}-\ref{baryons_mass2}. 
It can be found that the mass spectra we calculated are consistent with the results obtained using other theoretical methods.

\begin{table}[h]
	\centering
	\begin{ruledtabular}
		\caption{ Mass spectra (in MeV) of $QQn$ systems from the BO approximation (this work), relativistic quark model \cite{Ebert:2002ig}, lattice QCD \cite{Mathur:2018epb}, and quark potential models \cite{Roberts:2007ni,Karliner:2014gca}. }
		\label{baryons_mass1}
		\begin{tabular}{ccccccc}
			
			state & $J^P$ & our work &\cite{Ebert:2002ig} & \cite{Mathur:2018epb} & \cite{Roberts:2007ni}  &  \cite{Karliner:2014gca} \\
			\hline
			$\Xi_{cc}$ & $\frac{1}{2}^+$ & $3621.4$ & 3620 &  & 3676 & $3627\pm12$ \\
			
			$\Xi_{cc}^*$ & $\frac{3}{2}^+$ & 3687.4 & 3727 &  & 3753 & $3690\pm12$ \\
			
			$\Xi_{bb}$ & $\frac{1}{2}^+$ & 10252.5 & 10202 &  & 10340 & $10162\pm12$ \\
			
			$\Xi_{bb}^*$ & $\frac{3}{2}^+$ & 10272.3 & 10237 &  & 10367 & $10184\pm12$ \\
			
			$\Xi_{cb}$ & $\frac{1}{2}^+$ & 6933.0 & 6933 & 6945 & 7011 & $6914\pm13$ \\
			
			$\Xi_{cb}'$ & $\frac{1}{2}^+$ & 6961.0 & 6963 & 6966 & 7047 & $6933\pm12$ \\
			
			$\Xi_{cb}^*$ & $\frac{3}{2}^+$ & 6985.6 & 6980 & 6989 & 7074 & $6969\pm14$ \\
			
		\end{tabular}
	\end{ruledtabular}
\end{table}

\begin{table}[h]
	\centering
	\begin{ruledtabular}
		\caption{Mass spectra (in MeV) of $QQs$ systems from the BO approximation (this work), relativistic quark model \cite{Ebert:2002ig}, lattice QCD \cite{Mathur:2018epb}, and quark potential models \cite{Roberts:2007ni,Patel:2024mfn}. }
		\label{baryons_mass2}
		\begin{tabular}{ccccccc}
			
			state & $J^P$ & our work &  \cite{Ebert:2002ig} & \cite{Mathur:2018epb} &\cite{Roberts:2007ni}  & \cite{Patel:2024mfn} \\
			\hline
			$\Omega_{cc}$ & $\frac{1}{2}^+$ & 3784.1 & 3778 &  & 3815 & 3547.6 \\
			
			$\Omega_{cc}^*$ & $\frac{3}{2}^+$ & 3859.1 & 3872 &  & 3876 & 3619.1 \\
			
			$\Omega_{bb}$ & $\frac{1}{2}^+$ & 10420.5 & 10359 &  & 10454 & 10309.3 \\
			
			$\Omega_{bb}^*$ & $\frac{3}{2}^+$ & 10443.0 & 10389 &  & 10486 & 10328.1 \\
			
			$\Omega_{cb}$ & $\frac{1}{2}^+$ & 7098.0 & 7088 & 6994 & 7136 & 6931.9 \\
			
			$\Omega_{cb}'$ & $\frac{1}{2}^+$ & 7130.7 & 7116 & 7045 & 7165 &  \\
			
			$\Omega_{cb}^*$ & $\frac{3}{2}^+$ & 7156.8 & 7130 & 7056 & 7187 & 6971.5 \\
			
		\end{tabular}
	\end{ruledtabular}
\end{table}

\section{Doubly heavy tetraquarks \label{S4}}

Using the model parameters extracted previously, in this section we further systematically study the mass spectra and decays of the $S$-wave doubly heavy tetraquarks.
Physically, the quantum numbers of all $S$-wave tetraquark systems can be $J^P=0^+$, $1^+$, and $2^+$.
In addition, for a tetraquark system containing $\bar{u}$ or $\bar{d}$ quarks, it has a definite isospin quantum number.
In the diquark-antidiquark configuration, the color of the diquark (antidiquark) in the tetraquark system can be either $\bar{3}$ ($3$) or $6$ ($\bar{6}$).
For simplicity, we consider the two color singlet cases separately as follows.

\textit{$|(Q_1Q_2)^{\bar{3}}(\bar{q}_3\bar{q}_4)^{3}\rangle$ configuration}. 
For the doubly heavy tetraquark $Q_1Q_2\bar{q}_3\bar{q}_4$ system, the color features of the heavy diquark $(Q_1Q_2)^{\bar{3}}$ and the light antidiquark $(\bar{q}_3\bar{q}_4)^{3}$ satisfy $\bar{3}\otimes3\rightarrow1$.
Thus, the universal interaction term of the tetraquark system is expressed as
\begin{equation}
     V_t=V(\bm{x_A},\bm{x_B})+V(\bm{x_A},\bm{x_B};\bm{x_1},\bm{x_2}),
\end{equation}
with the heavy-heavy quark interaction
\begin{equation}
    V(\bm{x_A},\bm{x_B})=-\frac{2}{3}\frac{\alpha_s}{r_{AB}}+ V_{\text{conf}},
\end{equation}
and the light quark-related interactions
\begin{equation}
	\begin{aligned}
		V(\bm{x_A},\bm{x_B};\bm{x_1},\bm{x_2})=&-\frac{2}{3}\frac{\alpha_s^{\prime\prime}}{r_{12}}-\frac{1}{3}\frac{\alpha_s^\prime}{r_{A1}}-\frac{1}{3}\frac{\alpha_s^\prime}{r_{A2}}\\
		&-\frac{1}{3}\frac{\alpha_s^\prime}{r_{B1}}-\frac{1}{3}\frac{\alpha_s^\prime}{r_{B2}}.
	\end{aligned}
\end{equation}
In the above interaction parts, $\alpha_s^{\prime\prime}$ represents the strong coupling constant between two light antiquarks. 
$r_{12}$ represents the distance between two light antiquarks.
$r_{A1}$, $r_{A2}$, $r_{B1}$, and $r_{B2}$ represent the distances between heavy quarks and the light antiquarks.

Now we calculate the total ground-state energy $E$ of the $S$-wave doubly heavy tetraquark systems.
In the BO approximation, the two heavy quarks are treated as approximately stationary, while the light antiquarks move around them. 
Therefore, the interaction between light antiquarks is weaker than the heavy-heavy and heavy-light quark interactions.
In specific calculations, the light-light antiquark interaction can be regarded as a perturbative part.

Ignoring the interaction between two light antiquarks, a doubly heavy tetraquark system can be divided into two orbitals.
Similar to the doubly heavy baryon, these two orbitals are related to $\bar{q}_3$ ($Q_1\bar{q}_3$ and $Q_2\bar{q}_3$)  as well as $\bar{q}_4$ ($Q_1\bar{q}_4$ and $Q_2\bar{q}_4$) respectively.
Thus, the Hamiltonian for light antiquark $\bar{q}_3$ can be written as
\begin{equation}
	H_{\bar{q}_3}=-\frac{1}{2m_{\bar{q}_3}}\nabla^2-\frac{1}{3}\frac{\alpha_s^\prime}{r_{A1}}-\frac{1}{3}\frac{\alpha_s^\prime}{r_{B1}}.
	\label{H_tq}
\end{equation}
And the radial variation wave function of the system can be expressed as
\begin{equation}
	f_{\bar{q}_3}=\frac{R(r_{A1})+R(r_{B1})}{\sqrt{2(1+\mathcal{J})}}.
	\label{wf_tq}
\end{equation}
Combining Eqs.(\ref{H_tq}) and (\ref{wf_tq}), we calculate the expectation value of the light antiquark as
\begin{equation}
E_{\bar{q}_3} = \frac{\langle R(r_{A1}),H_{\bar{q}_3}R(r_{A1})\rangle+\langle R(r_{B1}),H_{\bar{q}_3}R(r_{A1})\rangle}{1+\mathcal{J}}, 
\end{equation}
with
\begin{equation}
	\begin{aligned}
		&\langle R(r_{B1}),H_{\bar{q}_3}R(r_{A1})\rangle=(\frac{A}{m_{\bar{q}_3}}-\frac{2}{3}\alpha_s^\prime)\mathcal{I}-\frac{A^2}{2m_{\bar{q}_3}}\mathcal{J}, \\
		&\langle R(r_{A1}),H_{\bar{q}_3}R(r_{A1})\rangle=\frac{A^2}{2m_{\bar{q}_3}}-\frac{A}{3}\alpha_s^\prime-\frac{1}{3}\alpha_s^\prime\mathcal{K}.
	\end{aligned}
\end{equation}
For the $Q_1\bar{q}_4$ and $Q_2\bar{q}_4$ orbitals, we use the same calculation method to obtain the corresponding wave function $f_{\bar{q}_4}$ and energy eigenvalue $E_{\bar{q}_4}$.
Again, using the calculus of variations, we can compute the variational parameter and the corresponding minimized energy eigenvalue.

Next, we calculate the contribution of the perturbative part.
The perturbative Hamiltonian for the doubly heavy tetraquark system reads
\begin{equation}
	H_{\text{pert}}=-\frac{2}{3}\frac{\alpha_s^{\prime\prime}}{r_{12}}.
\end{equation}
Since the motions of $\bar{q}_3$ and $\bar{q}_4$ can be regarded as approximately independent, the wave function of the two non-interacting orbitals is
\begin{equation}
		f= \frac{f(3,4)}{\sqrt{\mathcal{N}}} =\frac{f_{\bar{q}_3}f_{\bar{q}_4}}{\sqrt{\mathcal{N}}},
\end{equation}
where the normalization factor is
\begin{equation}
\mathcal{N}=\int^\infty_{-\infty}|f(3,4)|^2d\tau.
\end{equation}
Thus, the first-order perturbation energy can be calculated as follows
\begin{equation}
	\Delta E=\langle f|H_{\text{pert}}|f\rangle.
\end{equation}

The BO potential of the $|(Q_1Q_2)^{\bar{3}}(\bar{q}_3\bar{q}_4)^{3}\rangle$ configuration can be rewritten as
\begin{equation}
	V_{\text{BO}}(r_{AB})= -\frac{2}{3}\frac{\alpha_s}{r_{AB}}+ V_{\text{conf}} + E_{{l}},
	\label{Tetra_BO}
\end{equation}
where the energy of light antiquarks is
\begin{equation}
	 E_{{l}} = E_{\bar{q}_3} + E_{\bar{q}_4} + \Delta E.
\end{equation}
Substituting the BO potential Eq.(\ref{Tetra_BO}) into Schr{\"o}dinger Eq.(\ref{heavy_S}), we can calculate the total ground-state energy $E$ of the doubly heavy tetraquark system.
Taking into account the color-spin hyperfine interaction term, the mass spectra of the tetraquark states and the corresponding wave functions can be obtained.

\textit{$|(Q_1Q_2)^{6}(\bar{q}_3\bar{q}_4)^{\bar{6}}\rangle$ configuration}. 
This doubly heavy tetraquark configuration is completely similar to $\text{H}_2$ in the electromagnetic interaction.
The color coupling between the heavy diquark $(Q_1Q_2)^6$ and the light antidiquark $(\bar{q}_3\bar{q}_4)^{\bar{6}}$ is $6\otimes\bar{6}\rightarrow 1$.
Except for the different color factors in the color Coulomb interaction, the specific calculation method is similar to that used in the study of the tetraquark $|(Q_1Q_2)^{\bar{3}}(\bar{q}_3\bar{q}_4)^{3}\rangle$ configuration. 
We will not give the detailed calculation process here.

\subsection{The $cc\bar{n}\bar{n}$ system}

\begin{figure}[h]
	\centering
	\includegraphics[width=0.85\linewidth]{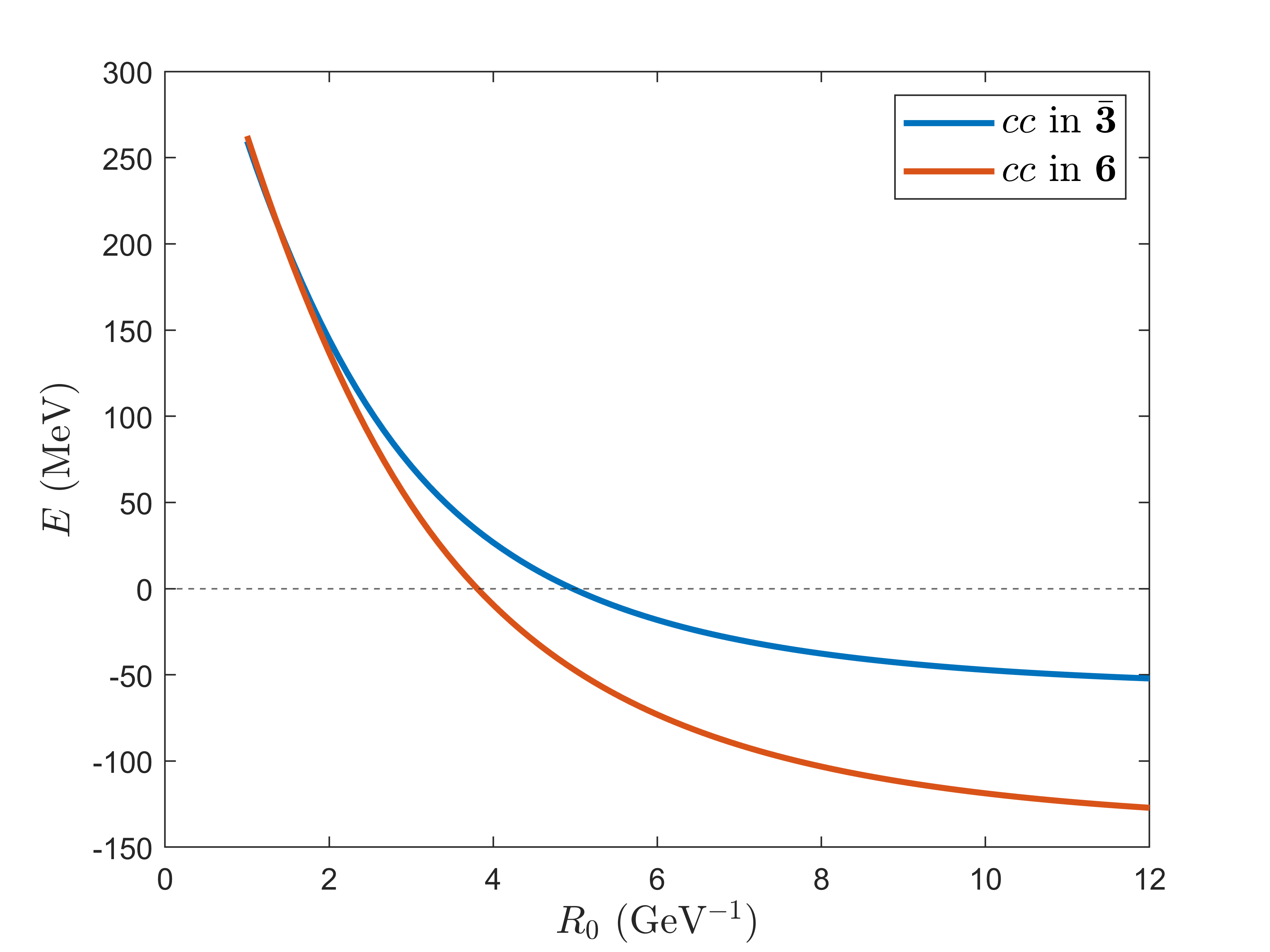}
	\caption{For the tetraquark $cc\bar{n}\bar{n}$ system, $R_{0}$-dependence of the ground-state energy $E$ for different colors.}
	\label{ccnn_ev}
\end{figure}

In the BO approximation, the total ground-state energies of the doubly charmed tetraquark $cc\bar{n}\bar{n}$ system have been calculated, as shown in Figure \ref{ccnn_ev}.
It can be found that no matter the color of the diquark $cc$ is $\bar{3}$ or $6$, the total ground-state energies of the tetarquark systems decrease with the increase of parameter $R_0$.
This shows that the introduction of linear potential strengthens the binding of the $cc\bar{n}\bar{n}$ system.
In addition, we can calculate the energy level splitting induced by the color-spin hyperfine interaction.
In the specific calculation, we give priority to studying the doubly charmed tetraquark $cc\bar{u}\bar{d}$ with isospin $I=0$.
For the diquark $cc$ in $\bar{3}$, the expectation value of the color-spin hyperfine interaction reads
\begin{equation}
	\langle H_{ss}\rangle = \frac{1}{2}(\kappa_{cc})^{\bar{3}}-\frac{3}{2}(\kappa_{nn})^{\bar{3}} = -150.8~\text{MeV}.
\end{equation}
For the diquark $cc$ in $6$, the expectation value of the color-spin hyperfine interaction is
\begin{equation}
	\langle H_{ss}\rangle = \frac{3}{4}(\kappa_{cc})^{\bar{3}}-\frac{1}{4}(\kappa_{nn})^{\bar{3}} = -20.2~\text{MeV}.
\end{equation}

\begin{figure}[htbp]
	\centering
	\begin{minipage}{0.244\textwidth}
		\centering
		\includegraphics[width=\linewidth]{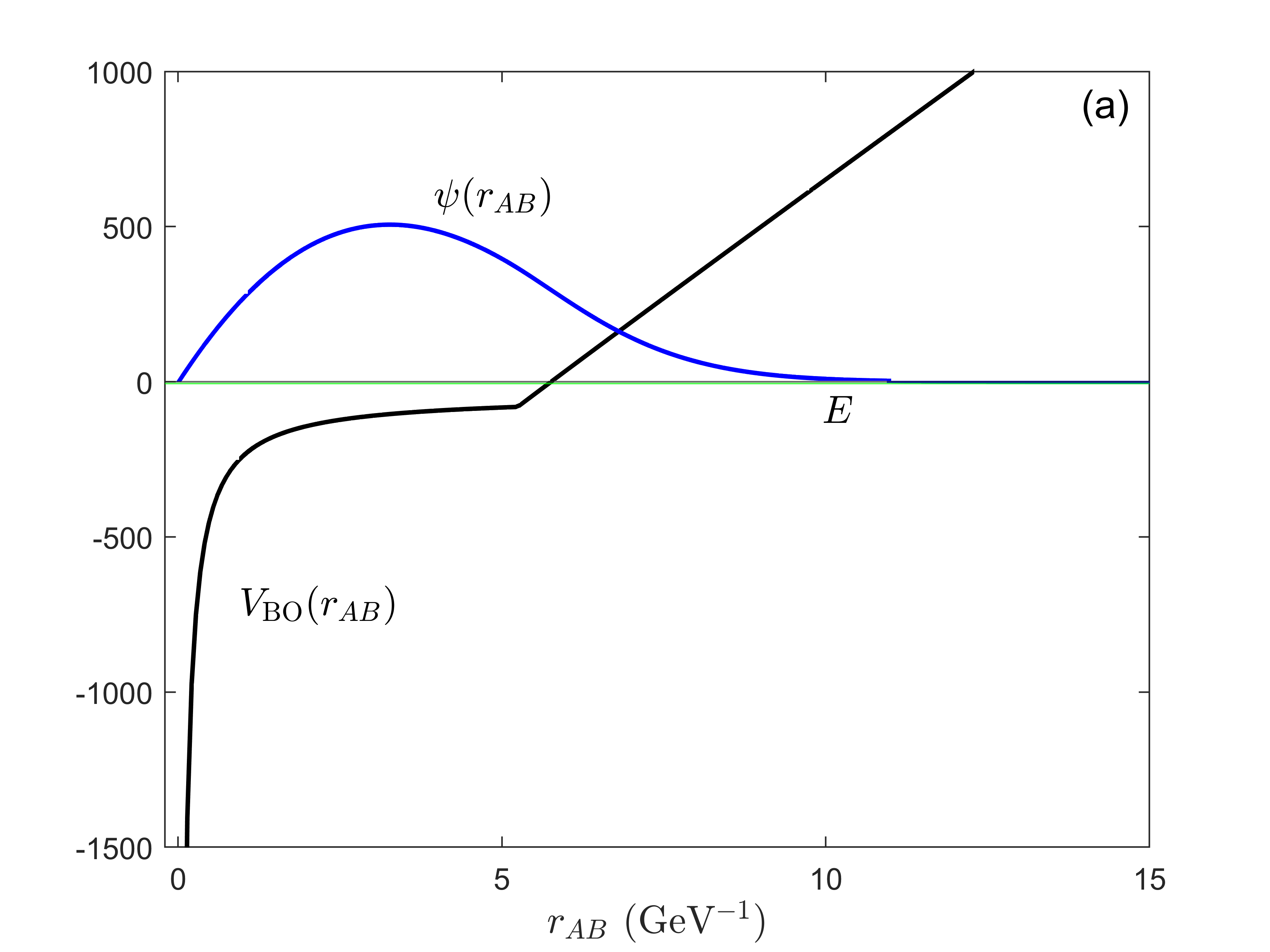}
	\end{minipage}
	\hspace{-0.5cm}
	\begin{minipage}{0.244\textwidth}
		\centering
		\includegraphics[width=\linewidth]{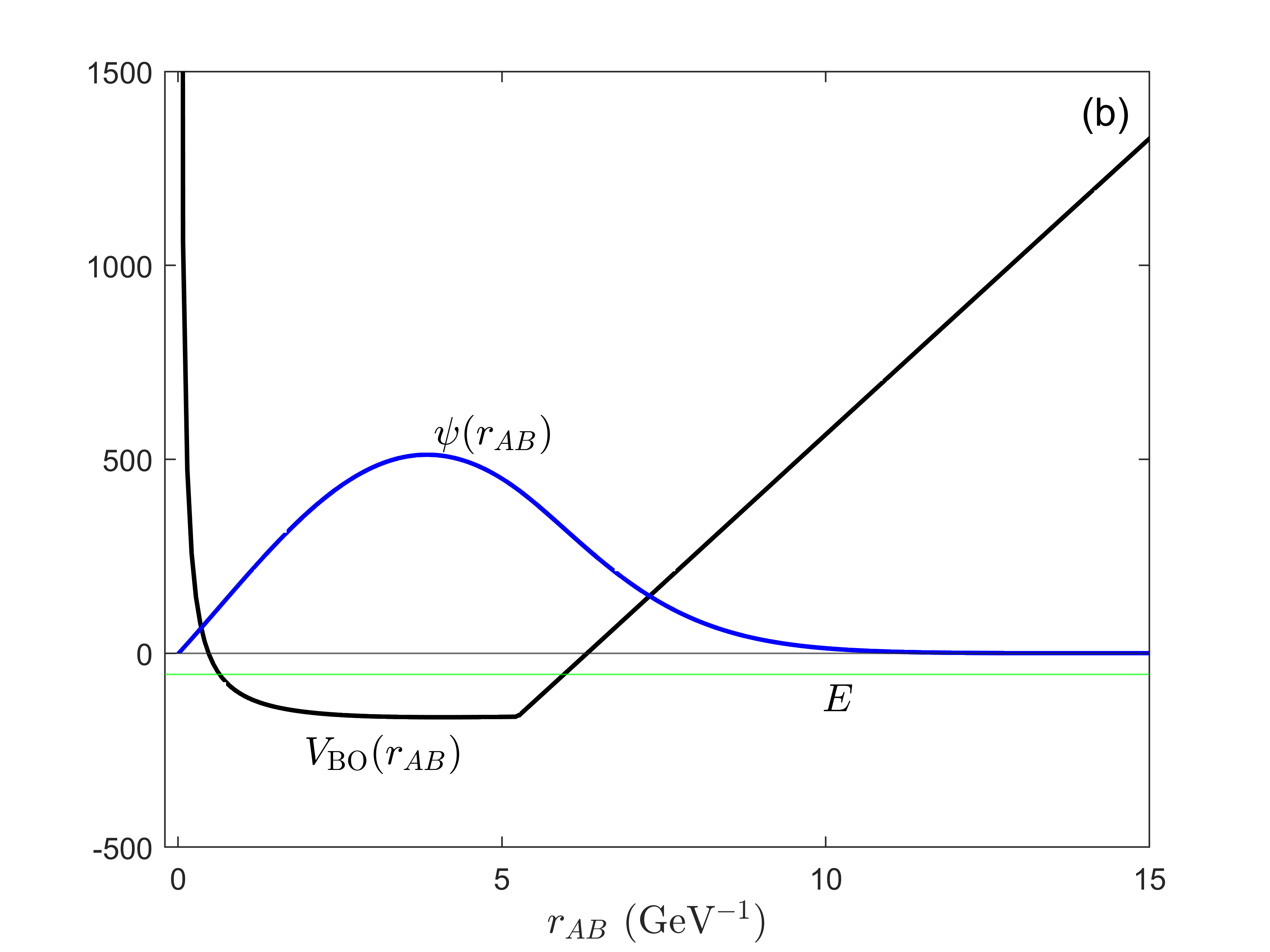}
	\end{minipage}
	
	\caption{For the tetraquark $cc\bar{u}\bar{d}$ system with quantum numbers $(I,J^P)=(0, 1^+)$, the color features of the heavy diquark are (a) $cc$ in $\bar{3}$, (b) $cc$ in $6$.
		The BO potential $V_{\text{BO}}(r_{AB})$, the total ground-state energy $E$ (in MeV), and the corresponding wave function $\psi(r_{AB})$ are shown.
	    In this calculation we take parameter $R_0 = 5.25$ GeV$^{-1}$. }
	\label{ccnn_potential}
\end{figure}

In the study of the doubly heavy tetraquark systems, the parameter $R_0$ can be obtained by fitting $T_{cc}^+$ and is directly used to predict other tetraquark states.
For the $S$-wave tetraquark $cc\bar{u}\bar{d}$ system, our calculation results show that the color coupling of the lowest state is $\bar{3}\otimes3\rightarrow1$.
Substituting the mass of $T_{cc}^+$ into Eq.(\ref{H}), we can obtain the total ground-state energy $E$ of the lowest state and the corresponding parameter $R_0$ as
\begin{equation}
	E = 75.6 \text{ MeV, }~ R_0=2.95 \text{ GeV}^{-1}.
	\label{Tcc_3}
\end{equation}
Obviously, in this case, the ground-state energy of the tetraquark $cc\bar{u}\bar{d}$ system is significantly greater than zero. 
This result implies that it is difficult for the tetraquark $cc\bar{u}\bar{d}$ system to form a bound state.

On the other hand, we calculated the case where the color coupling of the tetraquark $cc\bar{u}\bar{d}$ system is $6\otimes\bar{6}\rightarrow1$.
The calculated results are as follows
\begin{equation}
	E = -54.9 \text{ MeV, }~ R_0=5.25 \text{ GeV}^{-1}.
	\label{Tcc_6}
\end{equation}
Such a shallow tetraquark $cc\bar{u}\bar{d}$ bound state is consistent with our expectations.
And these features are in good agreement with the exotic hadron $T_{cc}^+$ recently observed by the LHCb collaboration.
Based on this, we suggest that $T_{cc}^+$ is more likely to be a good candidate for the tetraquark $|(cc)_0^{6}(\bar{u}\bar{d})_1^{\bar{6}}\rangle$ configuration with quantum numbers $(I,J^P)=(0, 1^+)$.
In Figure \ref{ccnn_potential}, we plot the BO potential, the total ground-state energy, and the corresponding wave function for different color couplings. 
The results show that when the color of the diquark $cc$ is $\bar{3}$, the BO potential mainly exhibits attraction. 
Correspondingly, when the color of the diquark $cc$ is $6$, the BO potential mainly exhibits repulsion, similar to the interaction between two protons in $\text{H}_2$.
Next, taken parameter $R_0=5.25$ GeV$^{-1}$, the ground-state energies of other $S$-wave doubly heavy tetraquarks can be systematically calculated, see Table \ref{dht_eigen}.
\begin{table}[htbp]
	\centering
	\begin{ruledtabular}
		\caption{In the BO approximation, the ground-state energies $E$ (in MeV) of other doubly heavy tetraquarks.
			     The superscripts ${\bar{3}}$ and ${6}$ correspond to the colors of the heavy diquark $Q_1Q_2$ in the tetraquark $Q_1Q_2\bar{q}_3\bar{q}_4$ system.
		         In this Table, we take the parameter $R_0 = 5.25$ GeV$^{-1}$.}
		\label{dht_eigen}
		\begin{tabular}{ccccccccc}
			  & $cc\bar{s}\bar{s}$ & $bb\bar{n}\bar{n}$ & $bb\bar{s}\bar{s}$  & $cc\bar{n}\bar{s}$ & $bb\bar{n}\bar{s}$ & $cb\bar{n}\bar{n}$ & $cb\bar{s}\bar{s}$ & $cb\bar{n}\bar{s}$ \\
			\hline
			$E^{\bar{3}}$  & -12.4 & -60.3 & -65.8 & -8.8 & -62.8 & -30.6 & -36.7 & -33.4 \\
			$E^{6}$        & -78.7 & -110.6 & -130.3 & -66.1 & -119.9 & -80.0 & -101.8 & -90.3 \\
		\end{tabular}
	\end{ruledtabular}
	
\end{table}

\begin{figure*}[htbp]
	\centering
	\begin{minipage}{0.333\textwidth}
		\centering
		\includegraphics[width=1.07\textwidth,height=0.86\textwidth]{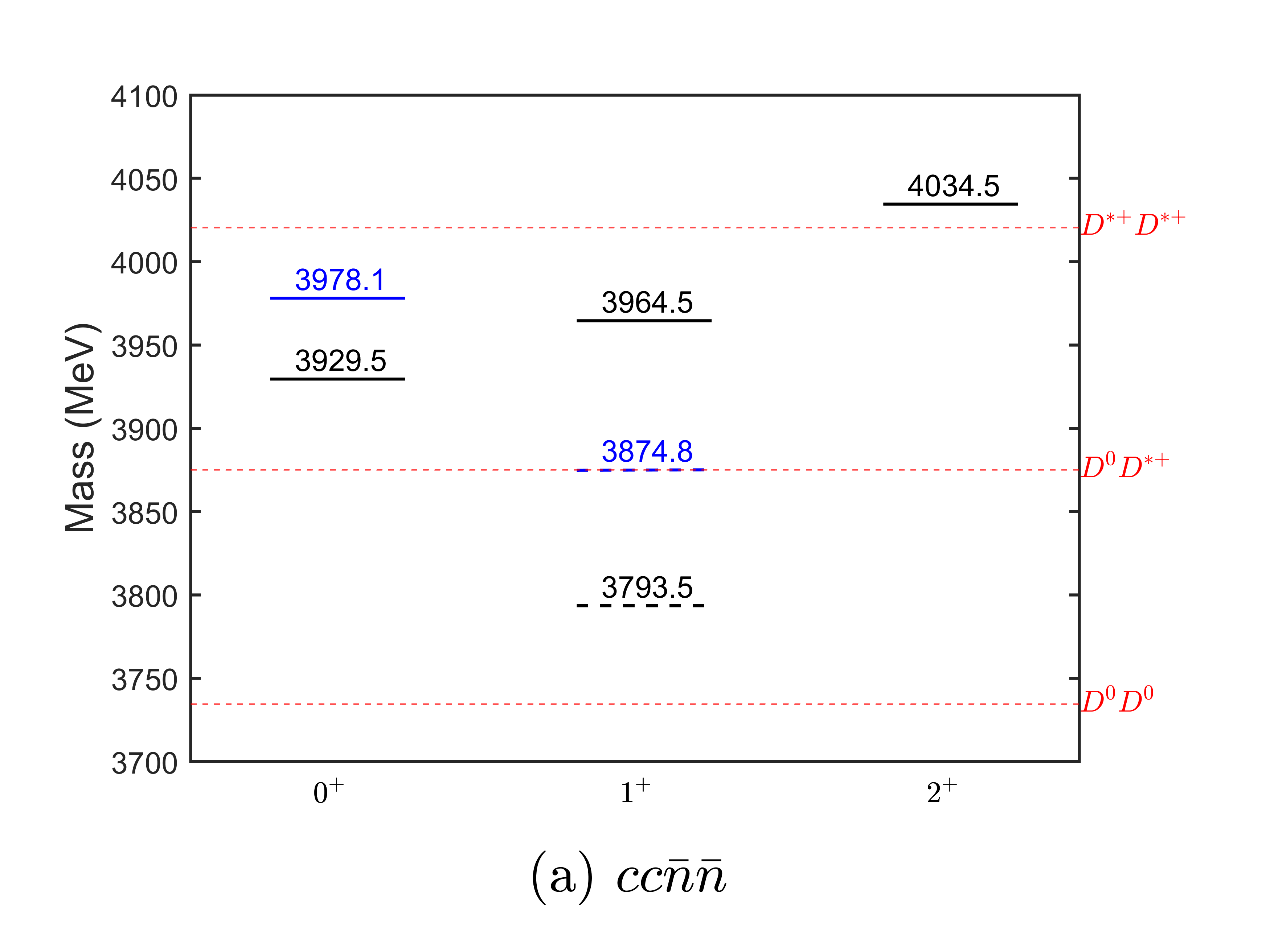}
	\end{minipage}\hfill
	\begin{minipage}{0.333\textwidth}
		\centering
		\includegraphics[width=1.07\textwidth,height=0.86\textwidth]{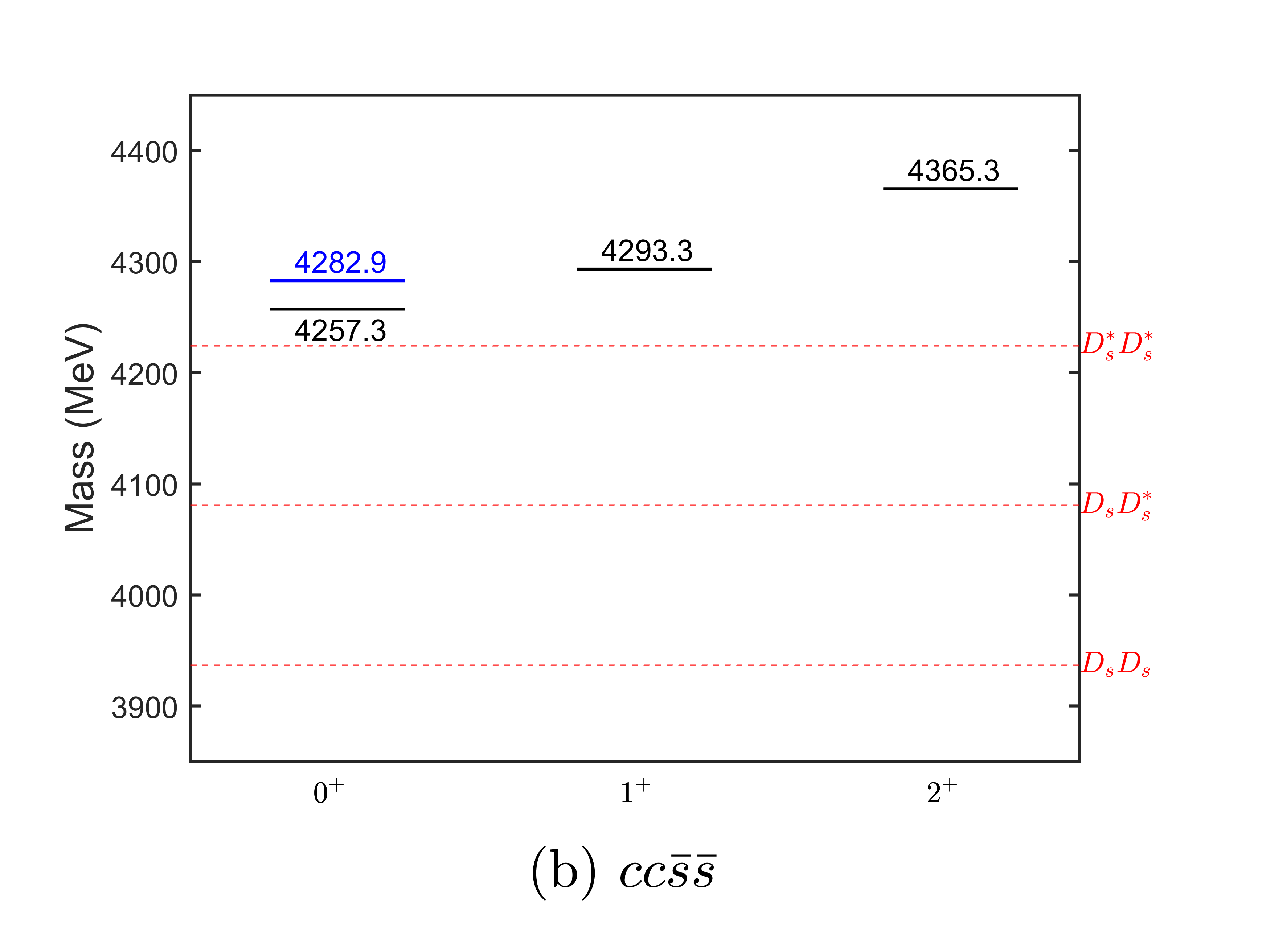}
	\end{minipage}\hfill
	\begin{minipage}{0.333\textwidth}
		\centering
		\includegraphics[width=1.07\textwidth,height=0.86\textwidth]{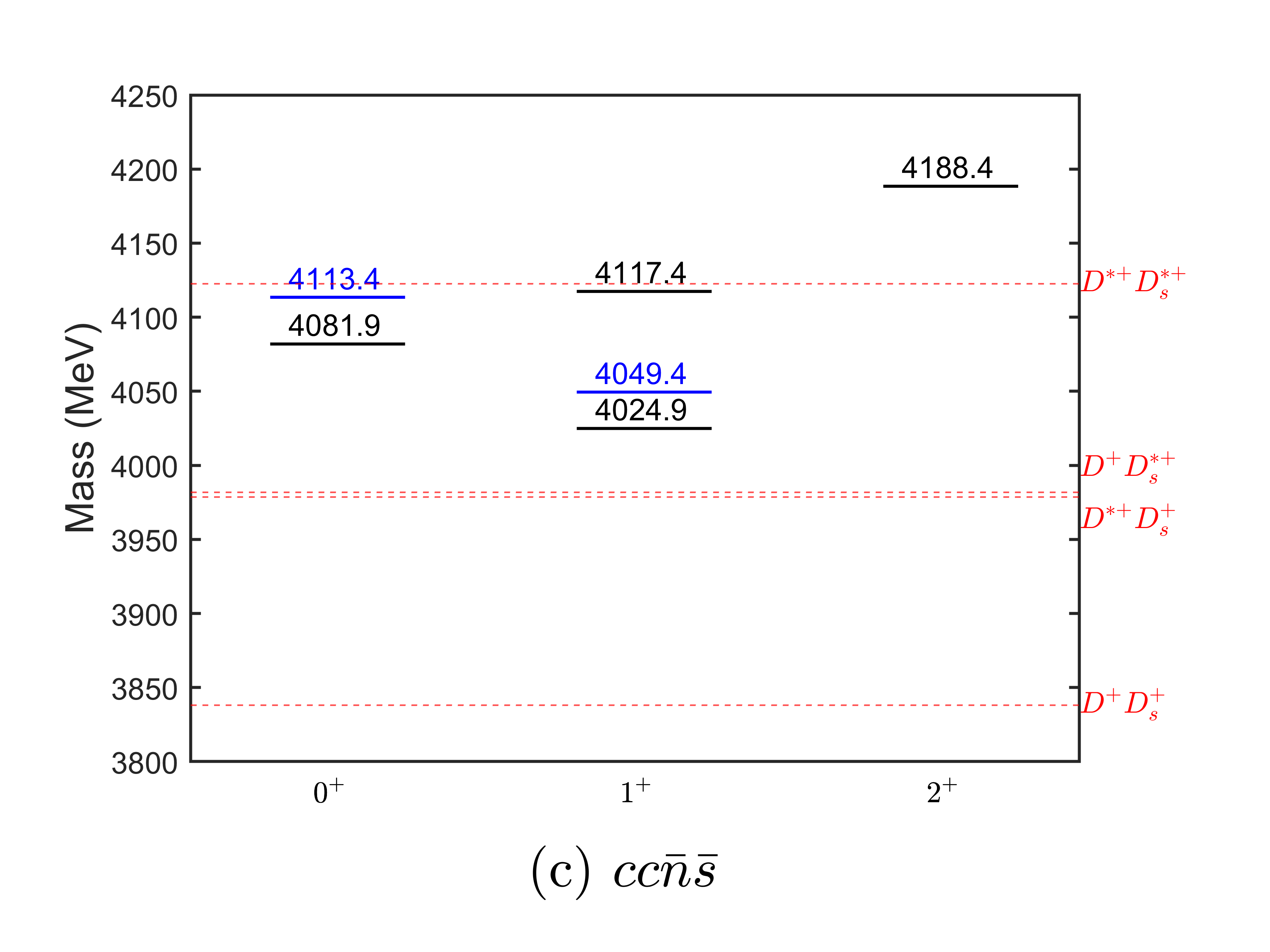}
	\end{minipage}
	
	
	\begin{minipage}{0.333\textwidth}
		\centering
		\includegraphics[width=1.07\textwidth,height=0.86\textwidth]{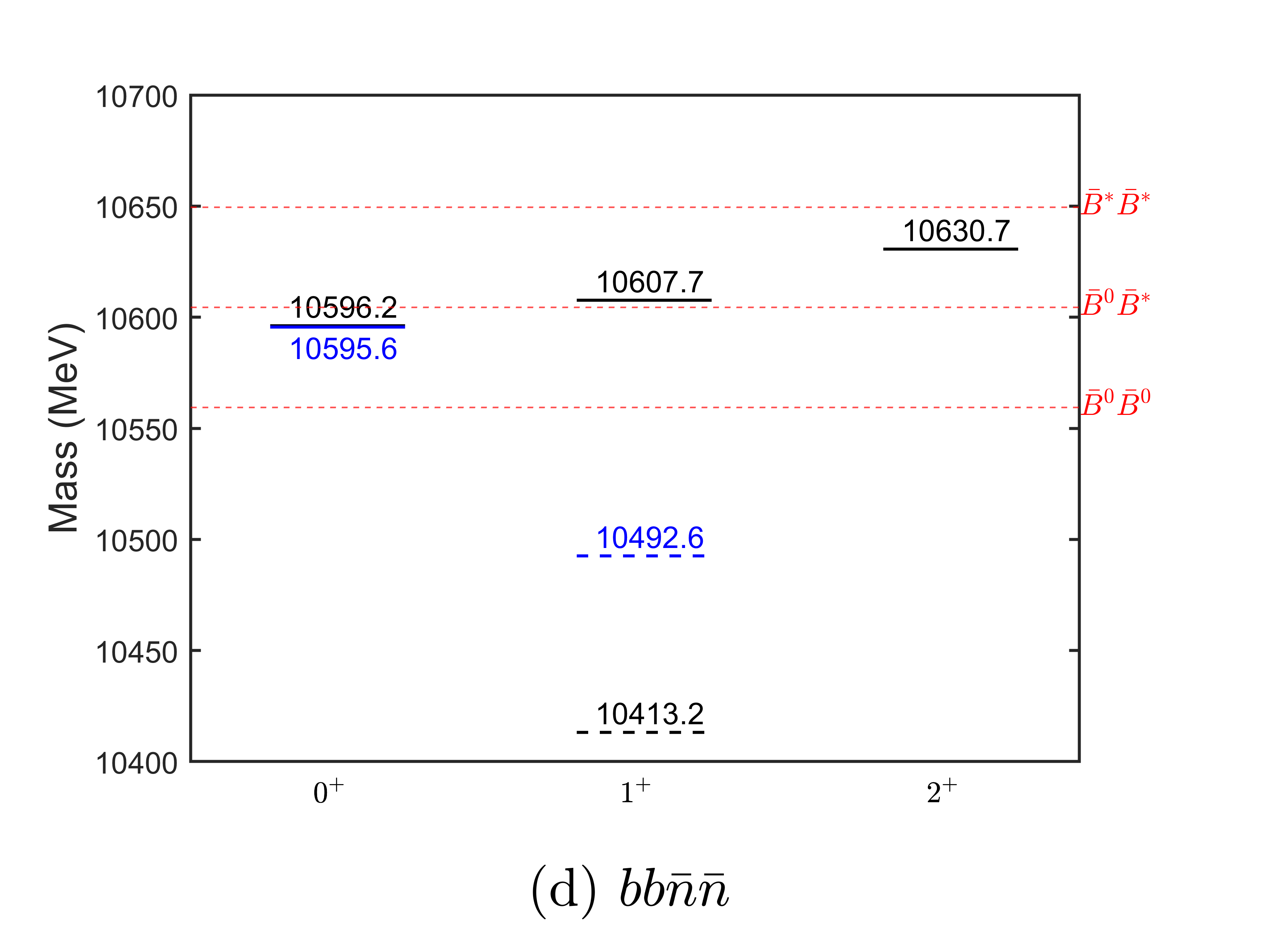}
	\end{minipage}\hfill
	\begin{minipage}{0.333\textwidth}
		\centering
		\includegraphics[width=1.07\textwidth,height=0.86\textwidth]{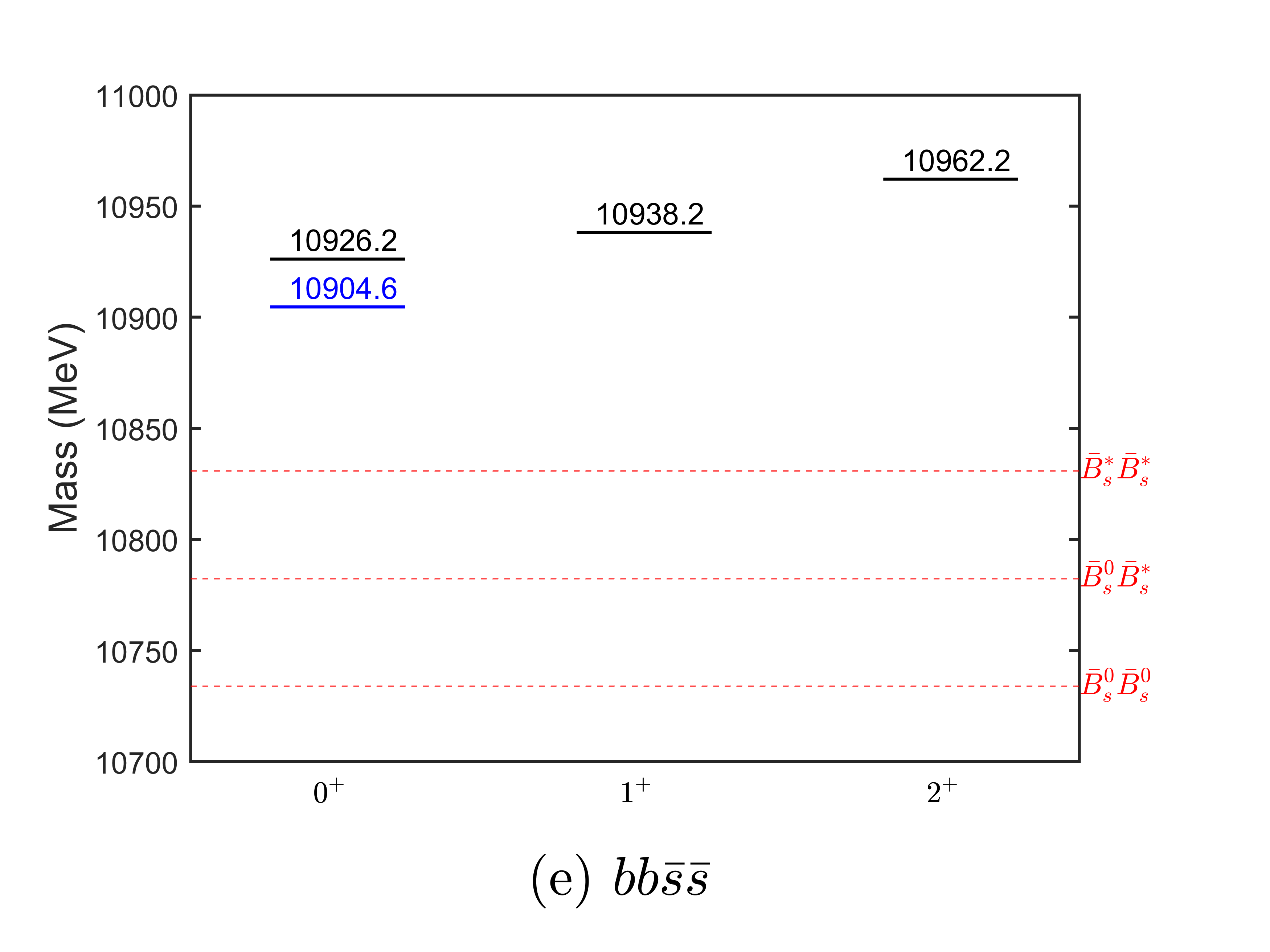}
	\end{minipage}\hfill
	\begin{minipage}{0.333\textwidth}
		\centering
		\includegraphics[width=1.07\textwidth,height=0.86\textwidth]{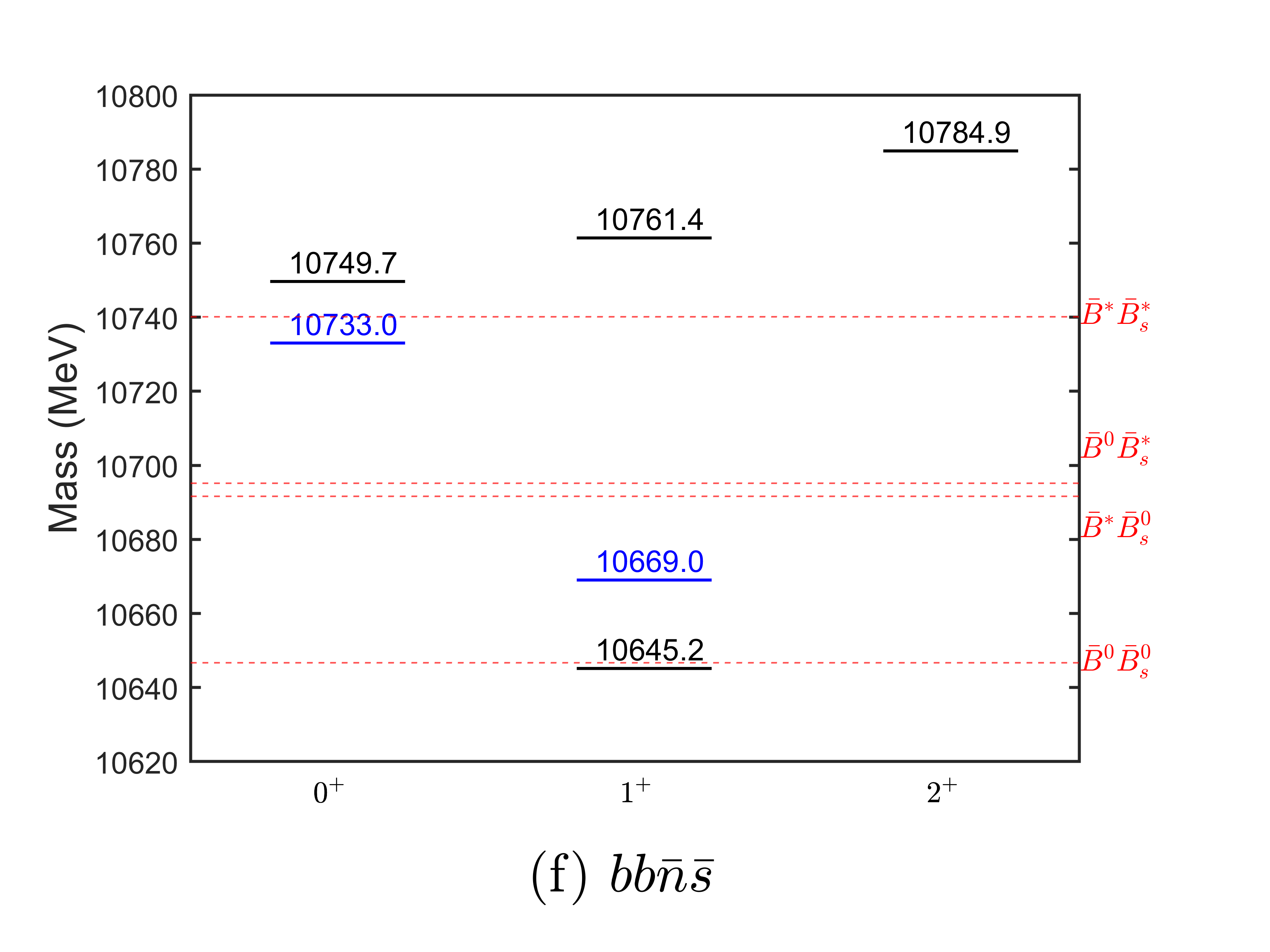}
	\end{minipage}\hfill
	
	
	\begin{minipage}{0.333\textwidth}
		\centering
		\includegraphics[width=1.07\textwidth,height=0.86\textwidth]{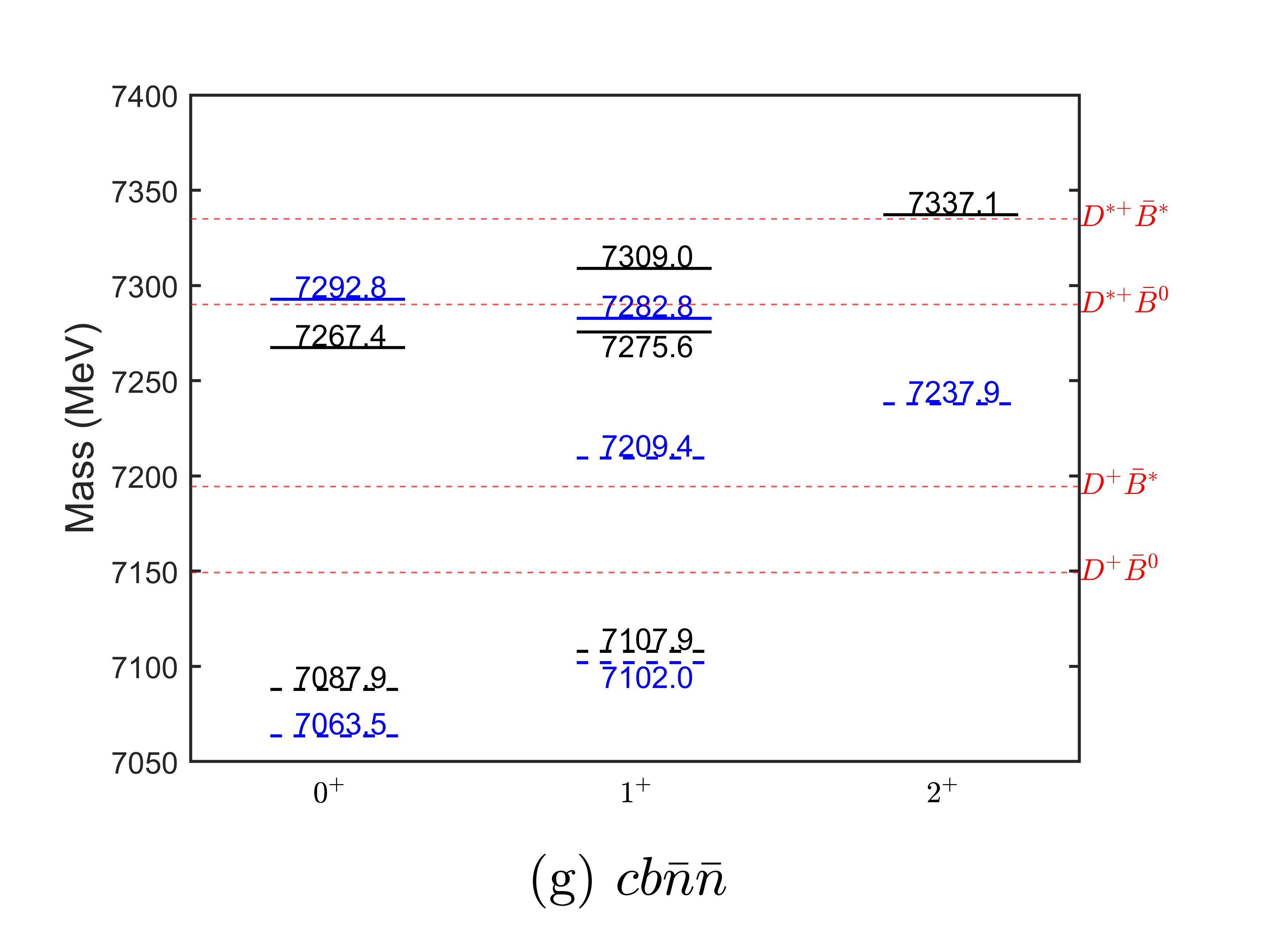}
	\end{minipage}\hfill
	\begin{minipage}{0.333\textwidth}
		\centering
		\includegraphics[width=1.07\textwidth,height=0.86\textwidth]{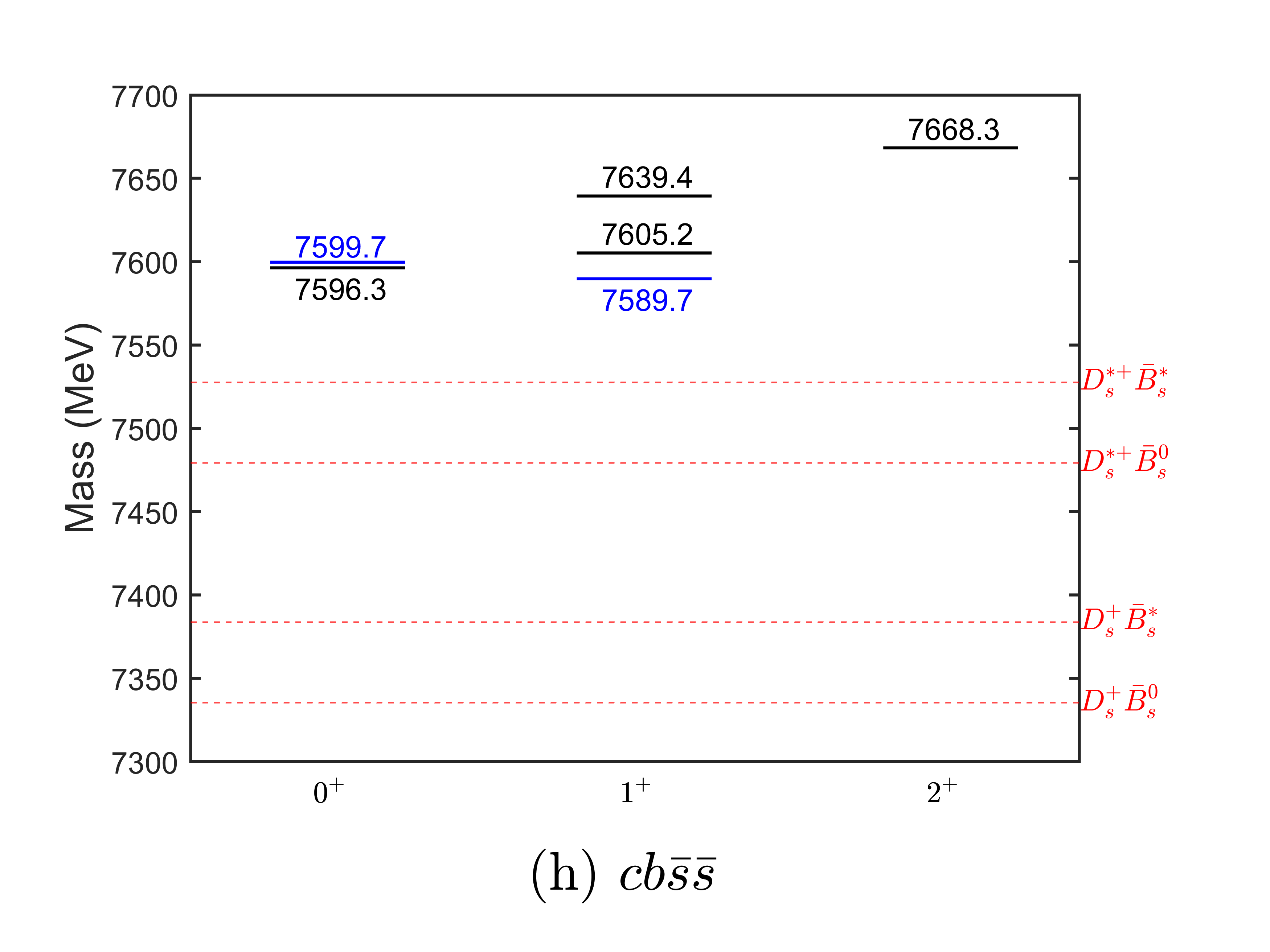}
	\end{minipage}\hfill
	\begin{minipage}{0.333\textwidth}
		\centering
		\includegraphics[width=1.07\textwidth,height=0.86\textwidth]{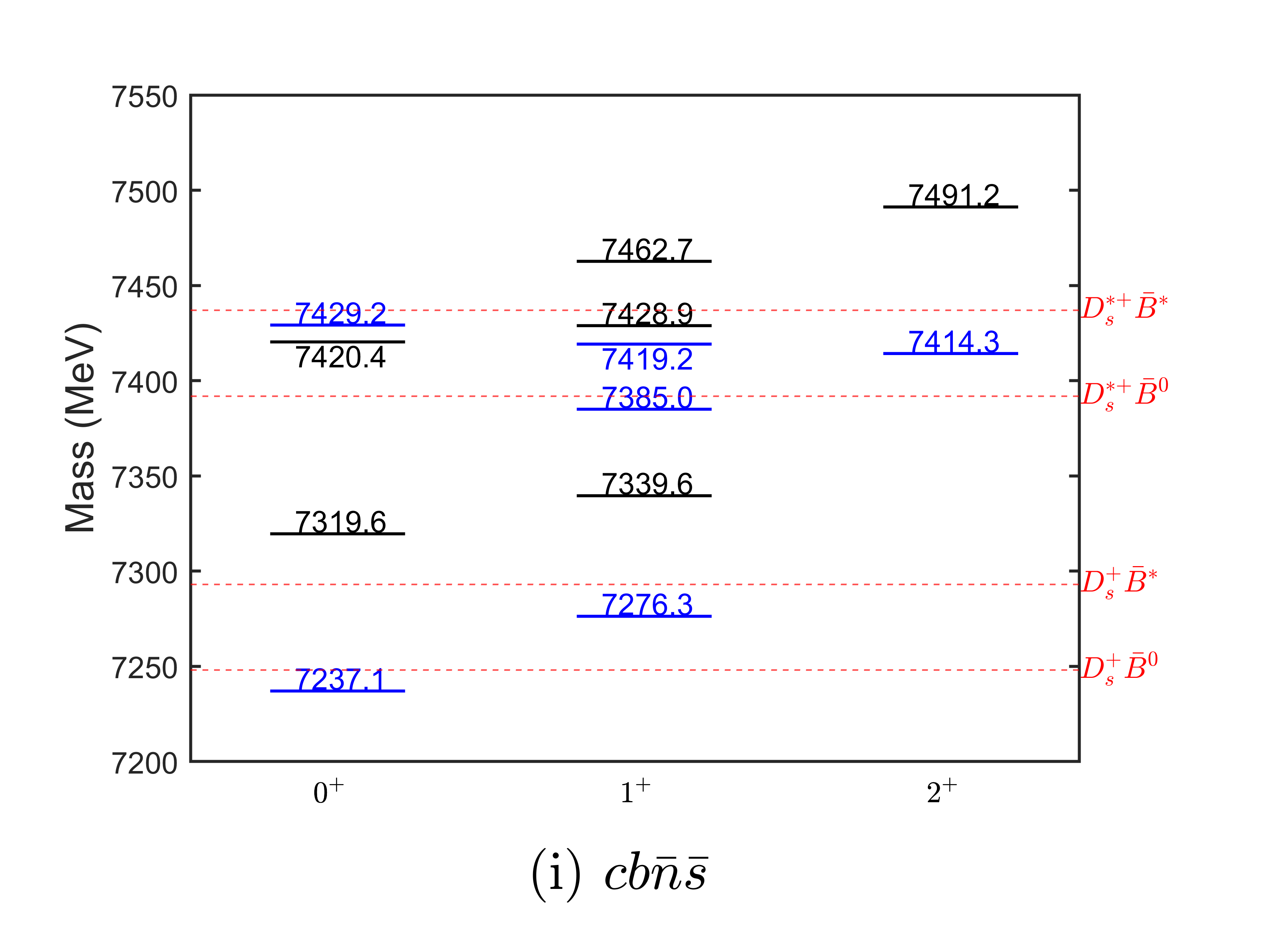}
	\end{minipage}
	
	\caption{Mass spectra of the $S$-wave doubly heavy tetraquark systems (a) $cc\bar{n}\bar{n}$, (b) $cc\bar{s}\bar{s}$, (c) $cc\bar{n}\bar{s}$, (d) $bb\bar{n}\bar{n}$, (e) $bb\bar{s}\bar{s}$, (f) $bb\bar{n}\bar{s}$ (g) $cb\bar{n}\bar{n}$, (h) $cb\bar{s}\bar{s}$, and (i) $cb\bar{n}\bar{s}$. 
		The blue mass spectra indicate that the heavy diquark color is in $6$. 
		The black mass spectra indicate that the heavy diquark color is in $\bar{3}$.
		The dashed (or solid) mass spectra in (a), (d), and (g) indicates isospin quantum number $I=0$ (or $1$).
		The red dashed lines are the corresponding meson-meson thresholds.}
	\label{mass_dht}
\end{figure*}

Now we can calculate the mass spectra of other $S$-wave tetraquark states in the $cc\bar{n}\bar{n}$ system, as shown in Figure \ref{mass_dht} (a).
To analyze the decay features of these tetraquark states, the corresponding meson-meson thresholds are also shown.
There are two $S$-wave tetraquark states with quantum number $J^P=0^+$. 
Their isospin quantum number $I=1$, and their quark components are $cc\bar{u}\bar{u}$ or $cc\bar{d}\bar{d}$.
Furthermore, the two tetraquark states lie between the $D^0D^0$ and $D^{*+}D^{*+}$ thresholds, allowing them to decay naturally into the two mesons $D^0D^0$ or $D^+D^+$.

For the doubly charmed tetraquark $cc\bar{n}\bar{n}$ system with quantum number $J^P=1^+$, there are three possible tetraquark states with masses of 3793.5 MeV, 3874.8 MeV, and 3964.5 MeV.
It is worth noting that 3793.5 MeV and 3874.8 MeV are isospin singlets, and the corresponding quark component is $cc\bar{u}\bar{d}$.
As mentioned before, 3874.8 MeV is slightly below the $D^0D^{*+}$ threshold and above the $D^0D^{0}\pi^+$ threshold.
Therefore, this state can decay naturally into the mesons $D^0D^{0}\pi^+$ via the off-shell process of the meson $D^{*+}$.
The lowest state at $3793.5$ MeV is significantly below the $D^0D^{*+}$ threshold, indicating that it may be a compact tetraquark state. 
If we only consider $S$-wave decays, this state cannot spontaneously decay into two mesons via the strong interaction processes.
The highest state, 3964.5 MeV, is an isospin triplet state. 
This state lies between the $D^0D^{*+}$ and $D^{*+}D^{*+}$ thresholds, and can spontaneously decay into mesons $D^0D^0$ or $D^+D^+$.

For the quantum number $J^P=1^+$, we only found a tetraquark state with a mass of 4034.5 MeV. 
Since the isospin quantum number of this state is $I=1$, its quark components may be the tetraquark $cc\bar{u}\bar{u}$ or $cc\bar{d}\bar{d}$.
In addition, this tetraquark state is above all corresponding meson-meson thresholds. 
According to the conservation of quantum numbers, the decay modes of this tetraquark state may be mesons $D^{*+}D^{*+}$ or $D^{*0}D^{*0}$.

\subsection{The $cc\bar{s}\bar{s}$ and $cc\bar{n}\bar{s}$ systems}  

In this subsection, the $S$-wave tetraquark states of the $cc\bar{s}\bar{s}$ and $cc\bar{n}\bar{s}$ systems are systematically investigated. 
For quantum numbers $J^P=0^+$, $1^+$, and $2^+$, the mass spectra of the tetraquark states and the corresponding meson-meson thresholds in these two systems are shown in Figures \ref{mass_dht} (b) and \ref{mass_dht} (c), respectively.

For the $cc\bar{s}\bar{s}$ system, it can be found from Figure \ref{mass_dht} (b) that there are four $S$-wave tetraquark states.
It is worth noting that these tetraquark states are above all thresholds. 
This indicates that all strong decay processes are allowed.
For example, the tetraquark states with quantum number $0^+$, 4257.3 MeV and 4282.9 MeV, can decay naturally into mesons $D_sD_s$ or $D_s^*D_s^*$. 
Similarly, the 4293.3 MeV with quantum number $1^+$ can decay into mesons $D_s D_s^*$ or $D_s^* D_s^*$, and the 4365.3 MeV with quantum number $2^+$ can decay into mesons $D_s^*D_s^*$.

For the $cc\bar{n}\bar{s}$ system, we found six $S$-wave tetraquark states, as shown in Figure \ref{mass_dht} (c).
There are two tetraquark states with quantum number $0^+$, and their masses are 4081.9 MeV and 4113.4 MeV respectively.
These two states are between the $D^+D_s^{+}$ and $D^{*+}D_s^{*+}$ thresholds, and can decay naturally into mesons $D^+D_s^{+}$.
For the quantum number $1^+$, the lowest two tetraquark states, 4024.9 MeV and 4049.4 MeV, are above the $D^{*+}D_s^+$ and $D^+D_s^{*+}$ thresholds, but significantly below the $D^{*+}D_s^{*+}$ threshold. 
This suggests that their main decay modes may be mesons $D^{*+}D_s^+$ or $D^+D_s^{*+}$. 
The remaining third state, 4117.4 MeV, is slightly below the $D^{*+}D_s^{*+}$ threshold, but above all other thresholds. 
This indicates that all possible strong decay processes in this state are possible.
In our calculations, there is only one tetraquark state with quantum number $2^+$, and its mass is 4188.4 MeV. 
This state is above all meson-meson thresholds, so it can naturally decay into mesons $D^{*+}D_s^{*+}$.

\subsection{The $bb\bar{q}\bar{q}$ and $cb\bar{q}\bar{q}$ systems}  

In the BO approximation theoretical framework, we further systematically calculated the low-lying $S$-wave tetraquark states of the doubly bottomed tetraquark $bb\bar{q}\bar{q}$ system and the mixed charm-bottom tetraquark $cb\bar{q}\bar{q}$ system. 
Flavor symmetries and color-spin interactions are also properly considered.
The calculated mass spectra are plotted in Figures \ref{mass_dht} (d)-(i).
To facilitate the analysis of the decay modes of these tetraquark states, we also plot the corresponding meson-meson thresholds in these figures.
The analysis is similar to that in the previous section, and we will not discuss each tetraquark state in detail here.
Notably, some $S$-wave stable tetraquark states below all corresponding meson-meson thresholds are predicted in the $bb\bar{q}\bar{q}$ and $cb\bar{q}\bar{q}$ systems, see Table \ref{tstable}.
In Breit–Wigner distribution, these tetraquark states are most likely very narrow peaks.
 
\begin{table}[htbp]
	\caption{\label{tstable} Some candidates for our predicted $S$-wave stable tetraquark states in the $bb\bar{q}\bar{q}$ and $cb\bar{q}\bar{q}$ systems.}
	\begin{ruledtabular}
		\begin{tabular}{lcc}
			system	& $IJ^P$ & mass (MeV) \\
			\hline			
			$bb\bar{n}\bar{n}$& $01^+$& 10413.2\\
			& $01^+$& 10492.6\\			
			$bb\bar{n}\bar{s}$& $\frac{1}{2}1^+$& 10645.2\\
			$cb\bar{n}\bar{n}$& $00^+$& 7063.5\\
			& $00^+$& 7087.9\\
			& $01^+$& 7102\\
			& $01^+$& 7107.9\\                    
			$cb\bar{n}\bar{s}$& $\frac{1}{2}0^+$& 7237.1\\                    
		\end{tabular}
	\end{ruledtabular}
\end{table}

\section{Summary \label{S5}}

In summary, we systematically study the hydrogen-like structures in the strong interaction based on the BO approximate theory. 
In this paper, these hydrogen-like structures refer specifically to the $S$-wave doubly heavy baryons and doubly heavy tetraquark states.
Assuming that the wave functions of light quarks and heavy quarks in a hadron can be treated separately, we introduce the BO potential and simplify the three-body or four-body Schr{\"o}dinger equation to the two-body Schr{\"o}dinger equation.
This makes it easy to calculate the total ground state energies of hadron systems.
At the same time, in order to study the energy level splitting induced by the color-spin interaction, we construct the basis vector wave functions of the $S$-wave doubly heavy baryons and doubly heavy tetraquarks.
The model parameters are obtained by fitting conventional hadrons and are directly used to predict other hadron spectra.

For the $S$-wave doubly heavy baryon systems with different quantum numbers $J^P={\frac{1}{2}}^+$ and ${\frac{3}{2}}^+$, we use the doubly charmed baryon $\Xi_{cc}^{++}$ as input and determine the parameter $R_0=4.12$ GeV$^{-1}$. 
Then we systematically study the mass spectra of other $S$-wave doubly heavy baryons. 
The calculated results are in good agreement with the predictions of the relativistic quark model, lattice QCD, and quark potential models.
This shows the reliability and robustness of using the BO approximation to study hydrogen-like molecular ions in the strong interaction.
For the $S$-wave doubly heavy tetraquark systems with different quantum numbers $J^P=0^+$, $1^+$, and $2^+$, we use $T_{cc}^+$ as input and determine the parameter $R_0=5.25$ GeV$^{-1}$. 
The results show that $T_{cc}^+$ is a good candidate for the compact tetraquark $|(cc)_0^{6}(\bar{u}\bar{d})_1^{\bar{6}}\rangle$ state with quantum number numbers $(I,J^P)=(0, 1^+)$.
Next, we systematically study other $S$-wave doubly heavy tetraquark states.
Among them, some stable narrow tetraquark states have been predicted and may be observed in future experiments.

Finally, In molecular physics and quantum chemistry, the BO approximation is one of the most successful theoretical methods to solve the problem of nucleon-electron interaction. 
In this work, we further introduce this method into the field of hadronic physics, hoping to provide a new approach for studying the hydrogen-like structures in the strong interaction.
Of course, the extraction of parameters is still model-dependent. 
We will further consider possible solutions in future work. 
At the same time, we also expect more theoretical and experimental studies on the hydrogen-like structures in the strong interaction.

\begin{acknowledgments}
	The authors would like to thank Xi Xia and Bowen Kang for helpful comments and discussions. This work was supported by the Scientific Research Foundation of Chengdu University of Technology under grant No.10912-KYQD2022-09557.
\end{acknowledgments}

\bibliography{boref}

\end{document}